%% file: bare_jrnl_compsoc.tex
\documentclass[10pt,journal,compsoc]{IEEEtran}
\usepackage{graphicx}
\usepackage{amssymb}
\usepackage{amsmath}
\usepackage{hyperref}
\usepackage{booktabs}
\usepackage{algorithm,algorithmicx,algpseudocode}
\usepackage{subfigure}
\usepackage{color}
\ifCLASSINFOpdf
\else
\fi
\begin{document}
%
\title{HyTasker: Hybrid Task Allocation in Mobile Crowd Sensing}
%
%
%
%

\author{Jiangtao Wang, Feng Wang, Yasha Wang, Leye Wang, Zhaopeng Qiu, Daqing Zhang, Bin Guo, Qin Lv
        
        
\IEEEcompsocitemizethanks{\IEEEcompsocthanksitem Jiangtao Wang, Feng Wang, Zhaopeng Qiu and Daqing Zhang are with School of EECS, Peking University, Beijing; Yasha Wang (corresponding author) is with National Engineering Research Center of Software Engineering, Peking University. All of them are with Key Laboratory of High Confidence Software Technologies, Ministry of Education.\protect\\
E-mail: jiangtaowang@pku.edu.cn;
\IEEEcompsocthanksitem Leye Wang is with Hong Kong University of Science and Technology.
\IEEEcompsocthanksitem Bin Guo is with Northwestern Polytechnical University.
\IEEEcompsocthanksitem Qin Lv is with University of Colorado Boulder.
}
\thanks{Manuscript received June 30, 2017.}}

%
%

\markboth{Journal of \LaTeX\ Class Files,~Vol.~14, No.~8, August~2015}%
{Shell \MakeLowercase{\textit{et al.}}: Bare Demo of IEEEtran.cls for Computer Society Journals}
%



\IEEEtitleabstractindextext{%
\begin{abstract}
Task allocation is a major challenge in Mobile Crowd Sensing (MCS). While previous task allocation approaches follow either the opportunistic or participatory mode, this paper proposes to integrate these two complementary modes in a two-phased hybrid framework called HyTasker. In the offline phase, a group of workers (called \textit{opportunistic workers}) are selected, and they complete MCS tasks during their daily routines (i.e., opportunistic mode). In the online phase, we assign another set of workers (called \textit{participatory workers}) and require them to move specifically to perform tasks that are not completed by the opportunistic workers (i.e., participatory mode).  Instead of considering these two phases separately, HyTasker jointly optimizes them with a total incentive budget constraint. In particular, when selecting opportunistic workers in the offline phase of HyTasker, we propose a novel algorithm that simultaneously considers the predicted task assignment for the participatory workers, in which the density and mobility of participatory workers are taken into account. Experiments on a real-world mobility dataset demonstrate that HyTasker outperforms other methods with more completed tasks under the same budget constraint.
\end{abstract}

}

\maketitle

\IEEEdisplaynontitleabstractindextext

%
\IEEEpeerreviewmaketitle

\IEEEraisesectionheading{\section{Introduction}\label{sec:introduction}}

\IEEEPARstart{W}{ith} the proliferation of sensor-rich mobile devices, a special form of crowdsourcing, called \textit{Mobile Crowd Sensing} (MCS) \cite{1,2}, has emerged as a new way of sensing and has drawn much attention from both academia \cite{3,4} and industry \cite{5,6}. MCS has stimulated in a variety of environmental, commercial, and social applications \cite{7,8,9,10,11,12,13}, where dynamically-moving citizens (called workers) contribute urban sensing information (e.g., traffic congestion status, air quality, and noise level) through mobile devices. Different from general online crowdsourcing, MCS requires workers' physical presence to perform location-dependent data collection tasks.

Task allocation or worker selection is one of the major challenges in MCS, which has a significant impact on the efficiency and quality of the sensing tasks \cite{14,17,12,19,25}. Recently, there have been many studies on MCS task allocation, such as \cite{20,21,22,23,24,28,29,33,35,36}, which can be divided into two categories based on the workers' movement patterns and participation mechanisms\cite{2,12,14,48}. (1) In the \textit{opportunistic mode}, an MCS system assigns tasks to a number of selected workers, who will complete the tasks during their daily routines without the need to change their routes \cite{20,21,22,23,24,28,29}. (2) In the \textit{participatory mode}, however, workers are required to change their original routes and move specifically to certain places to complete MCS tasks \cite{16,33,35,36}. Existing MCS solutions adopt either the opportunistic mode or the participatory mode to tackle the task allocation problem. However, both modes have their own advantages and disadvantages, which we elaborate as follows. 

The \textit{opportunistic mode} does not require knowledge of the workers' intended travel routes, so it is less intrusive for the workers and less costly for the task organizers. However, the sensing quality of the assigned tasks depends heavily on the workers' routine trajectories. For tasks that are located at places visited by few or even no workers, their sensing quality can be very poor. Additionally, in order to select an optimal set of workers, task allocation strategies based on the opportunistic mode usually need to predict the workers' trajectories, which significantly affects the optimality of the task allocation plan. Although different trajectory prediction algorithms \cite{15,24,35} have been proposed and proved to be effective to some extent, their accuracy cannot be theoretically guaranteed due to complicated and unpredictable real-life conditions.  Thus, the final sensing quality achieved for some tasks may be lower than expected.

The \textit{participatory mode} requires workers to move specifically to task locations, which can guarantee task completion. However, since workers need to deviate from their original routines and travel to task locations, it incurs extra travel cost and can be intrusive to the workers. It also increases the task organizers' incentive cost, since the task organizers usually have to pay extra incentive rewards to compensate for the traveling cost of the workers. Moreover, task allocation strategies based on the participatory mode only utilize mobile users who are willing to change their routes and travel intentionally for the tasks. As reported by some recent studies on human factors in MCS \cite{17,19}, a large proportion of mobile users are willing to contribute sensing data but do not want to change their routine trajectories. This group of mobile users is excluded from the candidate workers in the participatory mode, which is a waste of limited sensing resources.

In this paper, motivated by the complementary nature of these two modes, we propose a hybrid MCS task allocation framework, called HyTasker, which effectively integrates the opportunistic mode and the participatory mode via a two-phased design. In the offline phase, we recruit a number of workers (called \textit{opportunistic workers}) to complete sensing tasks during their routine trajectories.  In the online phase, we further assign some other workers (called \textit{participatory workers}) to locations where tasks cannot be completed by the opportunistic workers alone. Specifically, we study the typical budget-constrained MCS task allocation problem in this paper, where one task organizer launches a certain MCS campaign in the city with a number of location-dependent homogenous sensing tasks\cite{14,24,26,29,30}, with the goal of maximizing the number of completed tasks while keeping the total incentive rewards under a budget constraint.

Compared with the pure opportunistic or participatory mode, the advantage of HyTasker can be summarized in the following two aspects. 

First, \textit{from the perspective of the workers}, it naturally accommodates the workers' participation preferences and makes full use of the available human sensing resources. Although the workers may be willing to contribute sensing data for MCS campaigns, their preferred way of participation can be different. For example, some office employees are busy all day and do not have time to take a detour for task completion. In this case, they only accept to complete tasks on their daily routine trajectories. In contrast, some retired or unemployed citizens who have plenty of leisure time may be willing to move intentionally and complete tasks to earn incentive rewards. HyTasker assigns tasks based on the workers' preferences, hence making better use of the potential human sensing resources in the city.

Second, \textit{from the perspective of the task organizer}, HyTasker can achieve a better tradeoff between sensing quality and cost. Compared with pure participatory-mode approaches, it leverages some opportunistic workers to unintentionally complete tasks, which significantly reduces the incentive cost. In contrast to the pure opportunistic-mode approaches, it further improves the sensing quality by assigning some participatory workers to move and complete tasks in uncovered locations.

To illustrate how HyTasker works and further highlight the research challenges, we present a motivating example as follows. \textit{The city government launches an MCS campaign, called AirSense, for collecting real-time air quality information in different regions from 8:00am to 6:00pm every day in the city with a total budget constraint (e.g., 2000 USD per day). As the entire sensing area can be divided into 20 subareas (or called task locations), we can view the AirSense campaign as 20 homogenous sensing tasks in the same sensing period (e.g., from 8:00am to 6:00pm). 500 mobile users have registered as candidate workers in AirSense, and their historical records of connections to the cell towers are utilized by AirSense only for the purpose of task allocation after proper anonymization. According to their declared participation preferences, 350 workers are candidate opportunistic workers, while 150 are candidate participatory workers. As the budget is limited, AirSense cannot recruit all candidates to complete all tasks, and its goal is to design an effective task allocation mechanism to maximize the number of completed tasks. Hence, AirSense adopts the HyTasker framework for task allocation, which has the following two phases. First, in the offline phase, HyTasker selects a set of opportunistic workers from 350 candidates, and each of them is given a fixed reward (e.g., 10 USD per worker) for the entire sensing period \cite{21,24}. The selected opportunistic workers will collect sensing data for all tasks during their routine trajectories when they connect to the cell towers. Then, in the online phase (e.g., one hour before the end of sensing period), HyTasker spends the rest of the budget to assign another set of participatory workers from 150 candidates with uncompleted tasks so far. The participatory workers will change their daily routes to move intentionally for task completion, and the incentive reward each worker gets is in proportion to his/her travel distance (e.g., 2 USD per kilometer)} \cite{33,35}.

Given the basic design of HyTasker in the motivating example above, we can see the following research challenges. First, the two types of workers (i.e., opportunistic workers and participatory workers) share a total incentive budget, thus we cannot consider them separately. A na\"{\i}ve solution is to try different proportions of the budget division, and then directly adopt the state-of-the-art task allocation methods for opportunistic mode and participatory mode, respectively. However, as we cannot determine an optimal budget division plan, such a na\"{\i}ve solution may not perform well. Therefore, more sophisticated methods are needed to jointly optimize the offline and online phases. Second, in the offline phase of opportunistic worker selection, we need to consider possible online task assignments for the participatory workers in the future, which is challenging since we cannot foresee the precise locations of the participatory workers and the completion status of tasks. 

In an effort to address the objectives and challenges mentioned above, our work makes the following contributions:
\begin{enumerate}
	\item [(1)] By analyzing the complementary nature of the participatory-mode and opportunistic-mode MCS, we propose a two-phased hybrid task allocation framework, called HyTasker. It effectively integrates the two modes by selecting opportunistic workers in the offline phase and participatory workers in the online phase, via a joint optimization process. To the best of our knowledge, we are the first to combine these two modes in the MCS task allocation problem.
	\item [(2)] We propose a nested-loop greedy process to select the opportunistic workers by pre-considering future online task assignment of the participatory workers, which consists of two key mechanisms that are not adopted in the state-of-the-art MCS solutions. First, by considering the historical density of the participatory workers, HyTasker assigns higher priority to the opportunistic workers who are not only capable of completing more tasks but also can complete uncovered tasks in areas with fewer participatory workers. Second, it records each local-optimal subsets obtained during the greedy process, and further selects the optimal one by simultaneously predicting the task assignments of the participatory workers.
	\item [(3)] We evaluate HyTasker extensively using D4D \cite{39}, a real-world open dataset with 50,000 users' mobility traces. The experimental results demonstrate that HyTasker outperforms other methods with more completed tasks under the same budget constraint.
\end{enumerate}

\section{RELATED WORKS}
A number of research works exist for selecting MCS workers, who can complete MCS tasks during their daily routines without the need to change their original trajectories.  One group of studies considered the worker selection of a single MCS task with certain goals and constraints \cite{20,21,22,23,24, 26,28}. For example, the authors studied worker recruitment for a single MCS task, and they proposed different recruitment strategies to select a predefined number of workers so as to maximize the task's sensing quality \cite{20,21,22,23}, or select a minimum number of workers to ensure a certain level of sensing quality \cite{24,28}. Another group of studies attempted to optimize the overall utility of multiple concurrent sensing tasks in a multi-task-oriented MCS platform, where tasks share the limited resources \cite{29,30,31}. For example, both \cite{29} and \cite{30} proposed multi-task allocation algorithms to maximize overall system utility when the tasks share a limited incentive budget. The multi-task allocation strategy proposed in \cite{31} aims to optimize the overall utility when multiple tasks share a pool of workers with a sensing bandwidth constraint.

Another category of MCS tasks require workers to change their original routes and specifically move to certain places.  There are two models for task publishing, i.e., worker selected tasks (WST) \cite{32} and server assigned tasks (SAT) \cite{33, 35, 34, 36, 37}, in which tasks are selected by workers themselves or automatically assigned by the server, respectively. Our hybrid task allocation problem follows the SAT model. Prior studies in the SAT model \cite{33,35, 36, 37} assigned existing workers to tasks in the MCS system with various optimizing goals and constraints. For instance, the authors of \cite{33,35} aimed to maximize the number of completed tasks or overall task quality on the server side, while ensuring constraints on workers' maximum number of accepted tasks and task completion regions. The objective of \cite{37} is to minimize the traveling cost for completing a set of given tasks while seeking solutions that are socially fair.

The above studies adopt either opportunistic or participatory mode for MCS task allocation, while our work proposes a hybrid solution to achieve a better tradeoff between sensing quality and cost. Technically, our defined problem is more challenging, as we have to jointly optimize these two modes of task allocation with a shared incentive budget constraint. Different from existing MCS worker selection approaches, we develop a novel opportunistic worker selection algorithm with two unique mechanisms. First, instead of selecting workers who can only complete more tasks or cover larger areas \cite{21,24,30,29}, HyTasker assigns higher priority to those who can complete more tasks in participatory-worker-sparse areas. Second, while existing worker selection approaches commonly end when the total budget has been used up and output the final set of selected workers \cite{21,24,30,29}, HyTasker further records, estimates, and selects the best set of workers by simultaneously considering the predicted task assignments of the participatory workers.

\input{overview}
\input{core}
\input{eval}

\section{LIMITATION AND DISCUSSION}
This section discusses other issues that are not addressed in this work due to space and time constraints, which we plan to investigate in our future work.

		\textit{Different types of sensing tasks.} Although the current implementation of HyTasker is for environmental sensing tasks, the idea of the hybrid task allocation can be extended to other types of tasks if supported by the mobility data. As an example, for traffic status monitoring tasks, the subareas are divided based on the road sections rather than cell towers. Theoretically, if we use more precise trajectory data (e.g., GPS readings) to model more fine-grained mobility pattern of users (e.g., within the granularity of road sections), HyTasker can support the traffic status monitoring tasks. However, there are several challenges when extending HyTasker to such type of tasks in real-world settings, which can be added into our future work. For example, fine-grained localization raises the concerns of energy consumption and privacy leakage. How to balance the mobility prediction accuracy and these concerns is a challenging research question.

		\textit{Multiple heterogeneous tasks}. The current design of HyTasker focuses on MCS scenarios with homogeneous tasks. It would be useful to further study how HyTasker can be extended to a MCS platform with multiple heterogeneous sensing tasks. The challenges for such extension may include: (a) how to model the heterogeneity for different types of tasks, in terms of data quality, sensing period, sensing range, and sensing capability; (b) how to adopt the basic idea of HyTasker to optimize the overall utility of multiple heterogeneous tasks. With these challenges in mind, we will attempt to extend HyTasker to multi-task MCS scenarios in our future work.

\textit{Dynamic arrival of new tasks and workers}. In this paper, we assume that all tasks have been pre-published before the worker selection phase, so that the number of tasks and the distribution of task locations are already known and fixed. However, for a multi-task-oriented MCS platform, new tasks may be published anytime online. Moreover, new workers may also come to the platform continuously. Thus, how to tackle the dynamic arrival of new tasks and workers is a key challenge, which is not addressed in this paper. This challenge could lead to new research issues, such as how to predict the dynamic arrival of new tasks and workers, which will be added as our future work to extend the functionality of HyTasker. 

\textit{Timing for online task assignment}. In the online phase of HyTasker, we assign a set of participatory workers and require them  to move specifically to perform tasks that are not completed by the opportunistic workers. Here we assume that the participatory workers can complete the assigned tasks before the end time, and do not vary the timing of task assignment in this paper. In the experiments, we set this timing to one hour before the end of sensing period (e.g, 17:00 when the sensing period is 8:00-18:00). However, the best timing for online allocation still deserves further research in our future work. If the timing is too early, the participatory workers would complete more tasks which may be finished by opportunistic workers latter on, thus the total cost is higher. On the other hand, if the participatory workers are allocated too late, then they may not have enough time to travel and complete the tasks before the deadline. 

\textit{Different incentive models}. In HyTasker, the opportunistic workers get the same fixed reward for the entire sensing period, while the participatory workers get incentive rewards in proportion to their travel distances. In the MCS research community, there are actually a variety of incentive models \cite{42}, which are more complicated by considering multiple factors such as fairness, economic feasibility and data quality. In our future work, we plan to extend HyTasker by adopting more sophisticated incentive mechanisms (e.g., auction-based incentive models).

		\textit{Learning of task acceptance rate.}  In this work, we assume that the task acceptance rate of each participatory worker has already been learned, which is simulated in the experiments \cite{44,45,46}. To extend HyTasker to a wider range of application scenarios, we need to further improve it by learning and predicting the workers' task acceptance rate. For example, the authors in \cite{46} presented a learning framework based on workers' previous participation history with the consideration of incentive reward and task distance. However, several challenges exist to achieve a good prediction accuracy, which can be the direction of our future work. First, factors affecting users' decisions is very complex \cite{47} (e.g., task type, time availability, task distance, incentives, and even some emotional factors), and how to extract these features is non-trivial. Second, there are no historical participation records for new candidate workers, thus the prediction for their acceptance rate is challenging.

		\textit{Spatial correlation among tasks.} The goal of HyTasker is to maximize the number of completed tasks for a certain MCS campaign, in which we consider each task as equally important. Actually, the sensor readings in different subareas can be spatially correlated. For example, we can use the air quality information in one subarea to infer that of a nearby subarea. In our future work, we attempt to integrate mechanisms such as sparse crowd sensing \cite{28} into HyTasker to further reduce the sensing cost.

\section{CONCLUSION}
In this paper, we proposed a two-phased hybrid MCS task allocation framework, called HyTasker. In the offline phase, HyTasker selects a group opportunistic workers and requires them to complete MCS tasks during their daily routines. In the online phase, HyTasker assigns another set of participatory workers and requires them to move specifically to perform tasks that are not yet completed. Since the two types of workers share a total budget, we proposed a greedy based opportunistic worker selection process by simultaneously considering the predicted task assignments for the participatory workers. Experiments on a real-world mobility dataset show that HyTasker outperforms other baseline methods.

\ifCLASSOPTIONcaptionsoff
  \newpage
\fi



%

%

\begin{IEEEbiography}[{\includegraphics[width=1in,height=1.25in,clip,keepaspectratio]{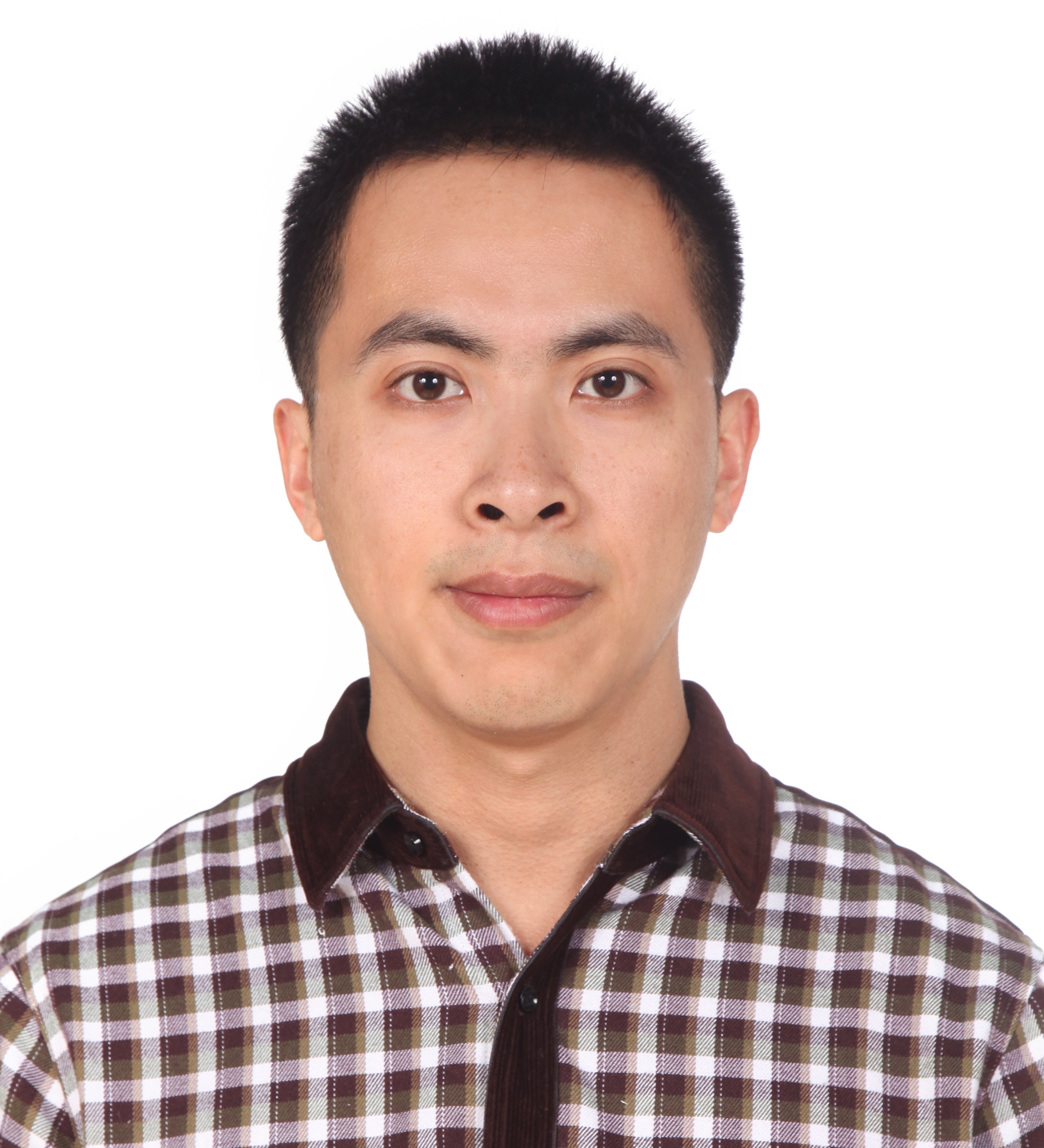}}]{Jiangtao Wang}
Jiangtao Wang received his Ph.D. degree in Peking University, Beijing, China, in 2015. He is currentlyan assistant professor in Institute of Software, School of Electronics Engineering and Computer Science, Peking University. His research interest includes collaborative sensing, mobile computing, and ubiquitous computing.
\end{IEEEbiography}

\begin{IEEEbiography}[{\includegraphics[width=1in,height=1.25in,clip,keepaspectratio]{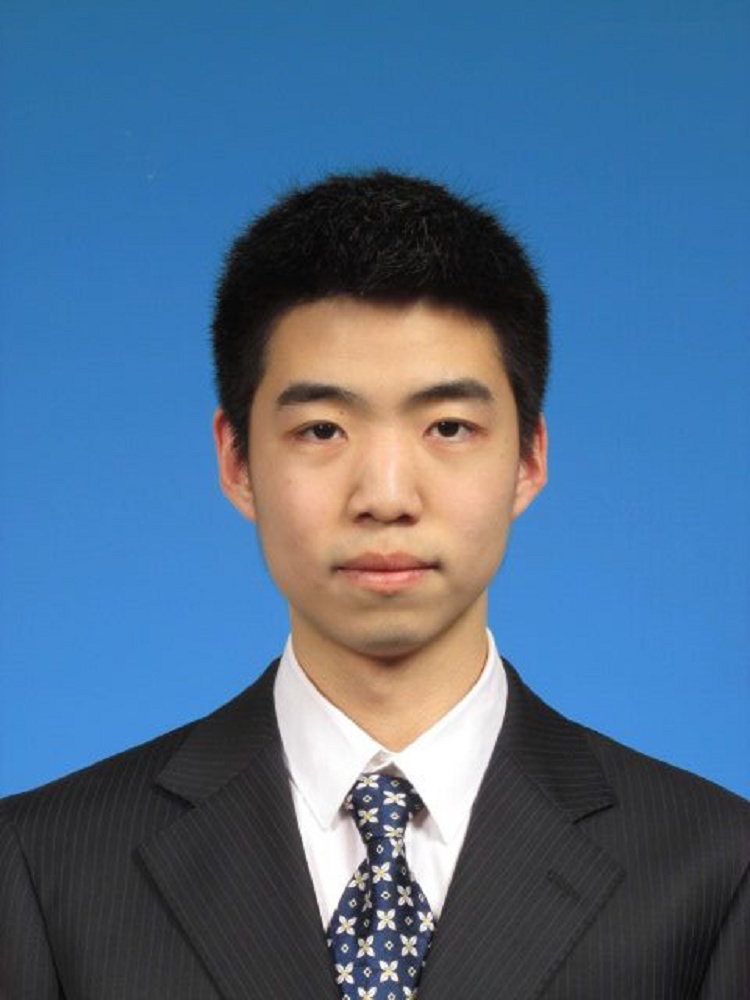}}]{Feng Wang}
	Feng Wang is an undergraduate student at School of Electronic Engineering and Computer Science, Peking University, China. His research interest is mobile crowd sensing.
\end{IEEEbiography}

\begin{IEEEbiography}[{\includegraphics[width=1in,height=1.25in,clip,keepaspectratio]{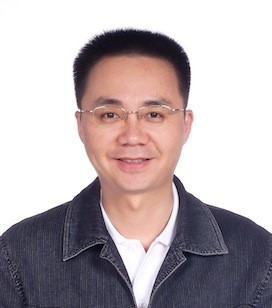}}]{Yasha Wang}
	Yasha Wang received his Ph.D. degree in Northeastern University, Shenyang, China, in 2003. He is a professor and associate director of National Research \& Engineering Center of Software Engineering in Peking University, China. His research interest includes urban data analytics, ubiquitous computing, software reuse, and online software development environment. He has published more than 50 papers in prestigious conferences and journals, such as ICWS, UbiComp, ICSP and etc. As a technical leader and manager, he has accomplished several key national projects on software engineering and smart cities. Cooperating with major smart-city solution providing companies, his research work has been adopted in more than 20 cities in China.
\end{IEEEbiography}

\begin{IEEEbiography}[{\includegraphics[width=1in,height=1.25in,clip,keepaspectratio]{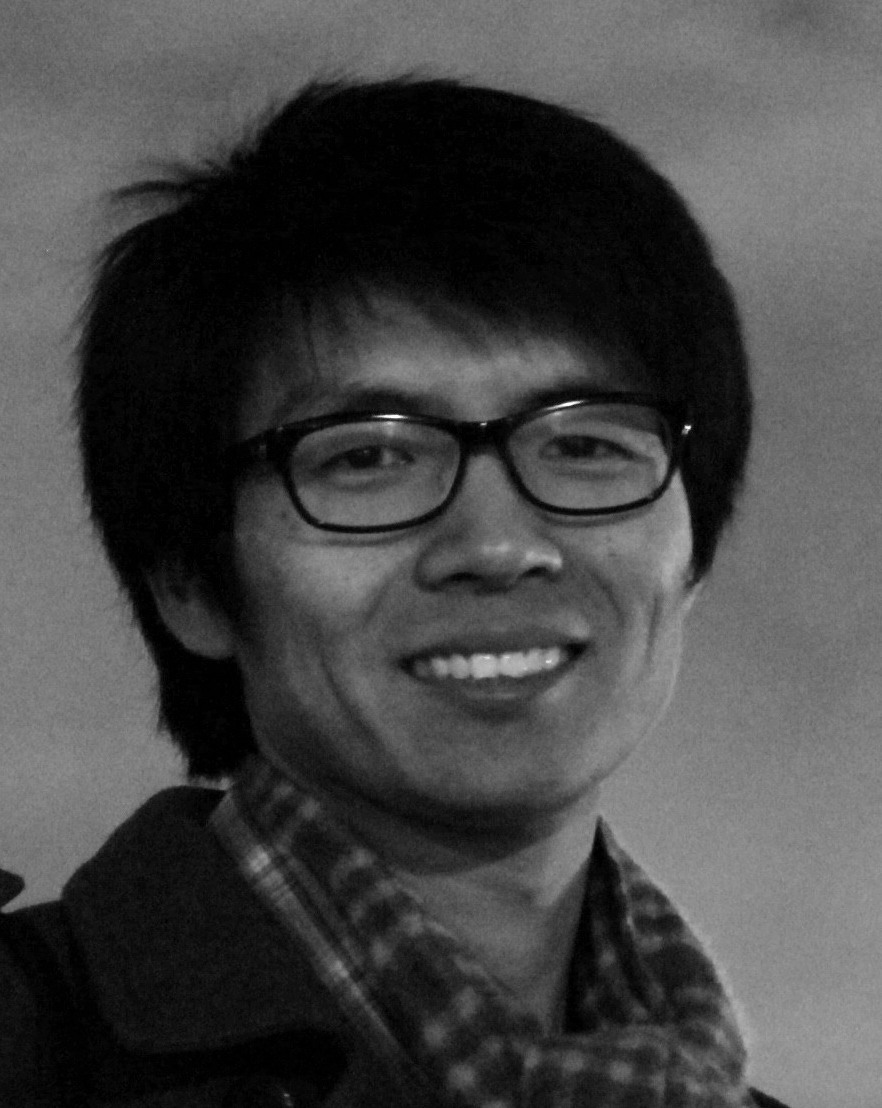}}]{Leye Wang}
	LEYE WANG obtained his Ph.D. from Institut Mines-T\'{e}l\'{e}com/T\'{e}l\'{e}com SudParis and Universit\'{e} Pierre et Marie Curie, France, in 2016. He received his M.Sc. and B.Sc. in computer science from Peking university, China. His research interests include mobile crowdsensing, social networks, and intelligent transportation systems.
\end{IEEEbiography}

\begin{IEEEbiography}[{\includegraphics[width=1in,height=1.25in,clip,keepaspectratio]{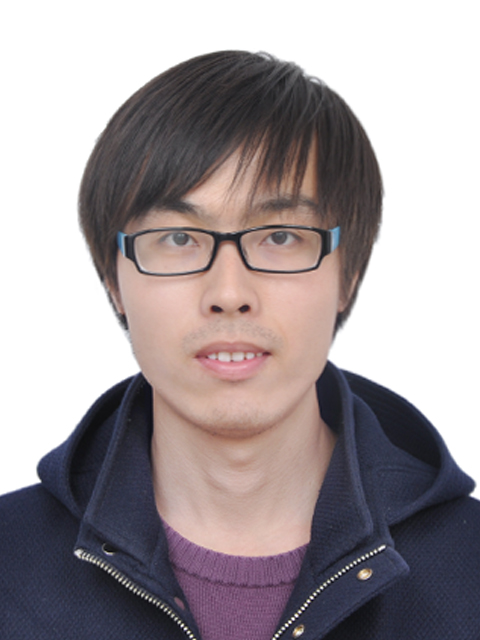}}]{Zhaopeng Qiu}
	Zhaopeng Qiu is an undergraduate student at School of Electronic Engineering and Computer Science, Peking University, China. His research interest is data analysis and NLP.
\end{IEEEbiography}

\begin{IEEEbiography}[{\includegraphics[width=1in,height=1.25in,clip,keepaspectratio]{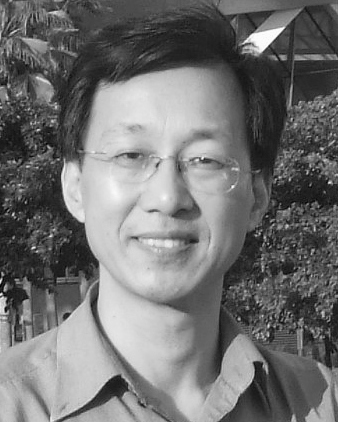}}]{Daqing Zhang}
Daqing Zhang is a professor at Peking University, China, and T\'{e}l\'{e}com SudParis, France. He obtained his Ph.D from the University of Rome “La Sapienza,” Italy, in 1996. His research interests include context-aware computing, urban computing, mobile computing, and so on. He served as the General or Program Chair for more than 10 international conferences. He is an Associate Editor for ACM Transactions on Intelligent Systems and Technology, IEEE Transactions on Big Data, and others.
\end{IEEEbiography}

\begin{IEEEbiography}[{\includegraphics[width=1in,height=1.25in,clip,keepaspectratio]{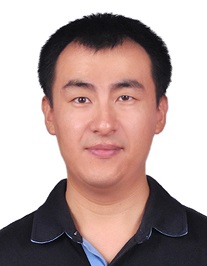}}]{Bin Guo}
Bin Guo received the PhD degree in computer science from Keio University, Japan, in 2009, and then was a postdoc researcher with Institut T\'{e}l\'{e}com SudParis, France. He is a professor with Northwestern Polytechnical University, China. His research interests include ubiquitous computing, mobile crowd sensing, and human-computer interaction.
\end{IEEEbiography}

\begin{IEEEbiography}[{\includegraphics[width=1in,height=1.25in,clip,keepaspectratio]{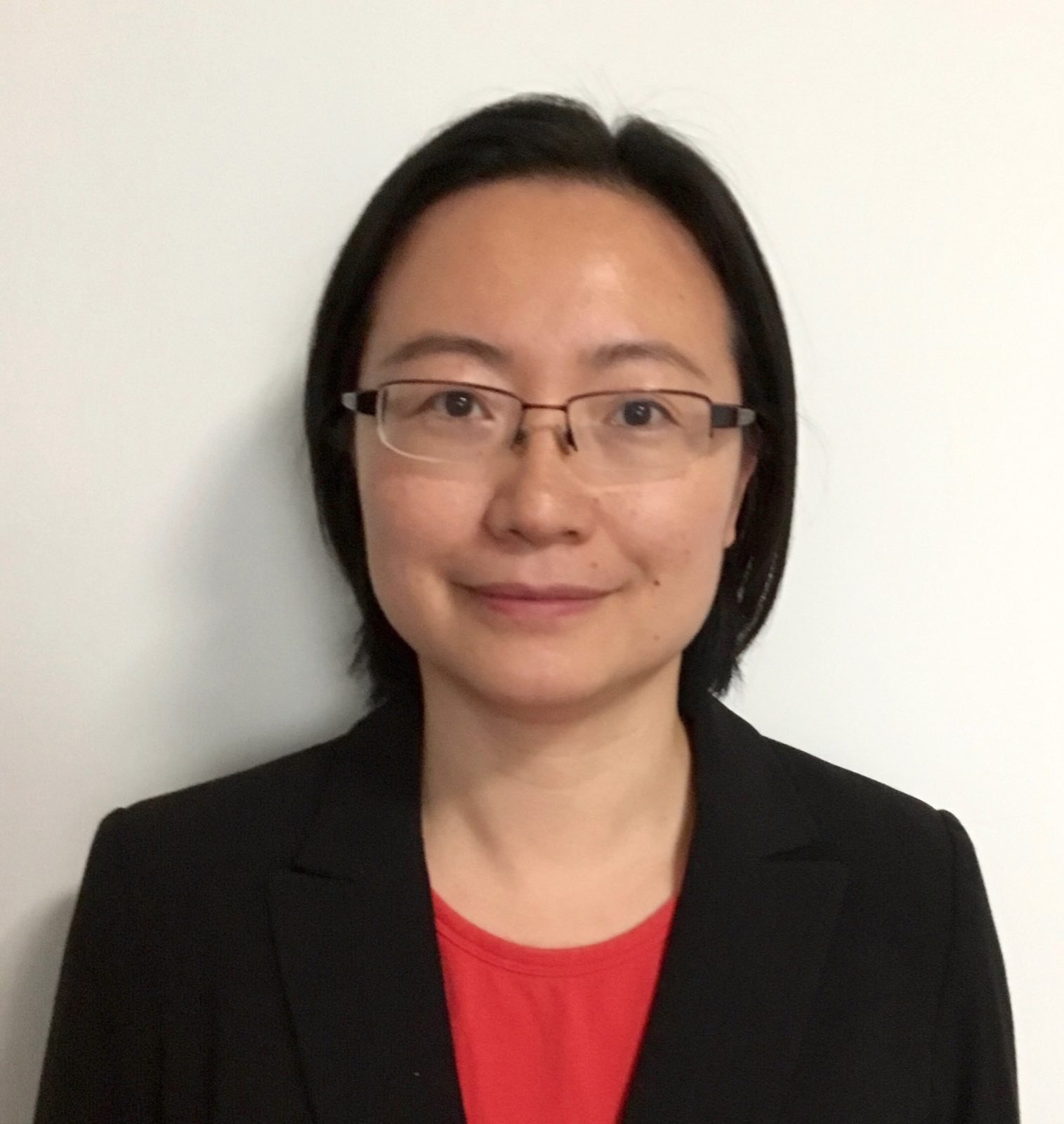}}]{Qin Lv}
	Qin Lv received her PhD degree in computer science from Princeton University in 2006. She is an associate professor in the Department of Computer Science, University of Colorado Boulder. Her main research interests are data-driven scientific discovery and ubiquitous computing. Her research spans the areas of multi-modal data fusion, spatial-temporal data analysis, anomaly detection, mobile computing, social networks, recommender systems, and data management. Her research is interdisciplinary in nature and interacts closely with a variety of application domains including environmental science, Earth sciences, renewable and sustainable energy, materials science, as well as the information needs in people’s daily lives. Lv is an associate editor of ACM IMWUT, and has served on the technical program committee and organizing committee of many conferences. Her work has more than 4,000 citation
\end{IEEEbiography}





\end{document}

%% file: overview.tex
\section{HYTASKER: SYSTEM OVERVIEW}
In this section, we first analyze and formulate the hybrid task allocation problem in HyTasker, and then describe the proposed HyTasker framework to solve this problem.

\subsection{Problem Analysis and Formulation}
Similar to some previous studies such as \cite{14, 24, 26}, this paper focuses on the task allocation of homogenous MCS tasks for urban environmental sensing. Specifically, the application scenario is that one organizer launches a certain type of MCS campaign (e.g., air quality sensing) during a certain period of time under a total budget constraint $B$. The entire sensing area can be divided into n subareas, and the data collection mission in each subarea is defined as a "task" in this paper. Thus, the MCS campaign consists of n location-dependent homogenous sensing tasks $T=\{t_1,t_2,...t_i...t_n\}$ during the same sensing period (e.g., 8:00am-6:pm). A task can be completed once a recruited worker moves into the corresponding subarea during the sensing period. The goal of HyTasker is to maximize the number of completed tasks by recruiting both opportunistic and participatory workers.

Similar to \cite{24,14, 26,31,25,30}, HyTasker also uses the cell tower as the sensing range of each subarea (i.e., sensing task), primarily due to two reasons: 1) The cell tower IDs of mobile phones are accessible in call logs, thus using cell towers as subarea division metrics can illustrate the core idea of HyTasker; 2) Many environment sensing tasks (e.g., air quality and temperature) can be carried out with a cell-tower level granularity. Hence, users with cell-tower positions can already conduct the tasks like air quality sensing well, and users do not need to upload their detailed GPS locations which are highly privacy-sensitive. Here, please note that the key insights and algorithms of HyTasker are not restricted by how the subareas are divided, and using cell towers as subarea division metrics is just an example to illustrate the core idea of HyTasker. If the subareas could be characterized more accurately, the proposed HyTasker framework could be easily adapted.

We divide the candidate workers into two disjoint categories $OW=\{ow_1,ow_2,...ow_j...ow_l\}$ and $PW=\{pw_1,pw_2,...pw_k...pw_m\}$ based on their self-defined preferences of participation mode. 1)\textit{The candidate opportunistic workers,} denoted as $OW$, would complete sensing tasks during their routine trajectories. 2)\textit{The candidate participatory workers.} This group of workers, denoted as $PW$, are more active and are willing to change their routes to complete sensing tasks. They actively report the participation information to the cloud server, including their current online location $P=\{p_1,p_2,...p_k...p_m\}$, spatial region $RW=\{rw_1,rw_2,...rw_k...rw_m\}$ within which they are willing to travel. Each of them can get incentive rewards in proportion to the actual travel distance online, which is denoted as $I*Distance(pw_k,PT_k)$, where $I$ is the proportion and $Distance(pw_k,PT_k)$ is the length of the shortest path for $pw_k$ to complete assigned task set $PT_k$. The maximum number of tasks $pw_k$ can be assigned is denoted as $L$.

 Based on the problem analysis above, we formally define the hybrid task allocation problem as follows. 1)\textit{In the offline phase}, our objective is to select a subset of candidate opportunistic workers $OW_f\subseteq OW$ and pay each of them a fixed and equal incentive reward $I_c$. Similar to \cite{24,14, 26,31}, HyTasker adopts a piggyback MCS paradigm for the opportunistic workers, in which the workers will complete sensing tasks online if and when they are connected to corresponding cell towers. The selected opportunistic workers will complete sensing tasks if they move into the task's geographic sensing range during the sensing time period. 2)\textit{In the online phase}, at timestamp $TS$, we aim to assign the participatory workers some tasks that are not completed by the opportunistic workers so far. We denote the full set of task-and-worker pairs as $V=PW\times T=\{(pw_k,t_i)|pw_k\in PW,t_i\in T\}$. So the objective of online task assignment is to select a subset $V_f$ of $V$ subject to the constraints (i.e., spatial region and maximum number of assigned tasks). Here, similar to previous studies such as \cite{44,45,46}, we assume that once a participatory worker $pw_k$ is assigned a task, the probability (called task acceptance rate) that s/he will accept the task is $ac_k$, which has already been learned from his/her previous participation history.

It is important to note that the online phase and the offline phase are correlated, because they share a total budget constraint. Specifically, for the MCS platform, the optimization goal is to maximize the total number of completed task set, denoted as $T_c$($T_c\subseteq T$), while keeping the total incentive reward under the budget constraint. The optimization problem can be formulated as follows:
\begin{equation}
Maximize\ |T_c|
\end{equation}
\begin{equation}
Subject\ to: I_c*|OW_f|+ \sum_{pw_k\in PW}I*Distance(pw_k, PT_k)\leq B
\end{equation}

\subsection{HyTasker Overview Design}
\begin{figure}[!t]
	
	\centering \includegraphics[width=.5\textwidth]{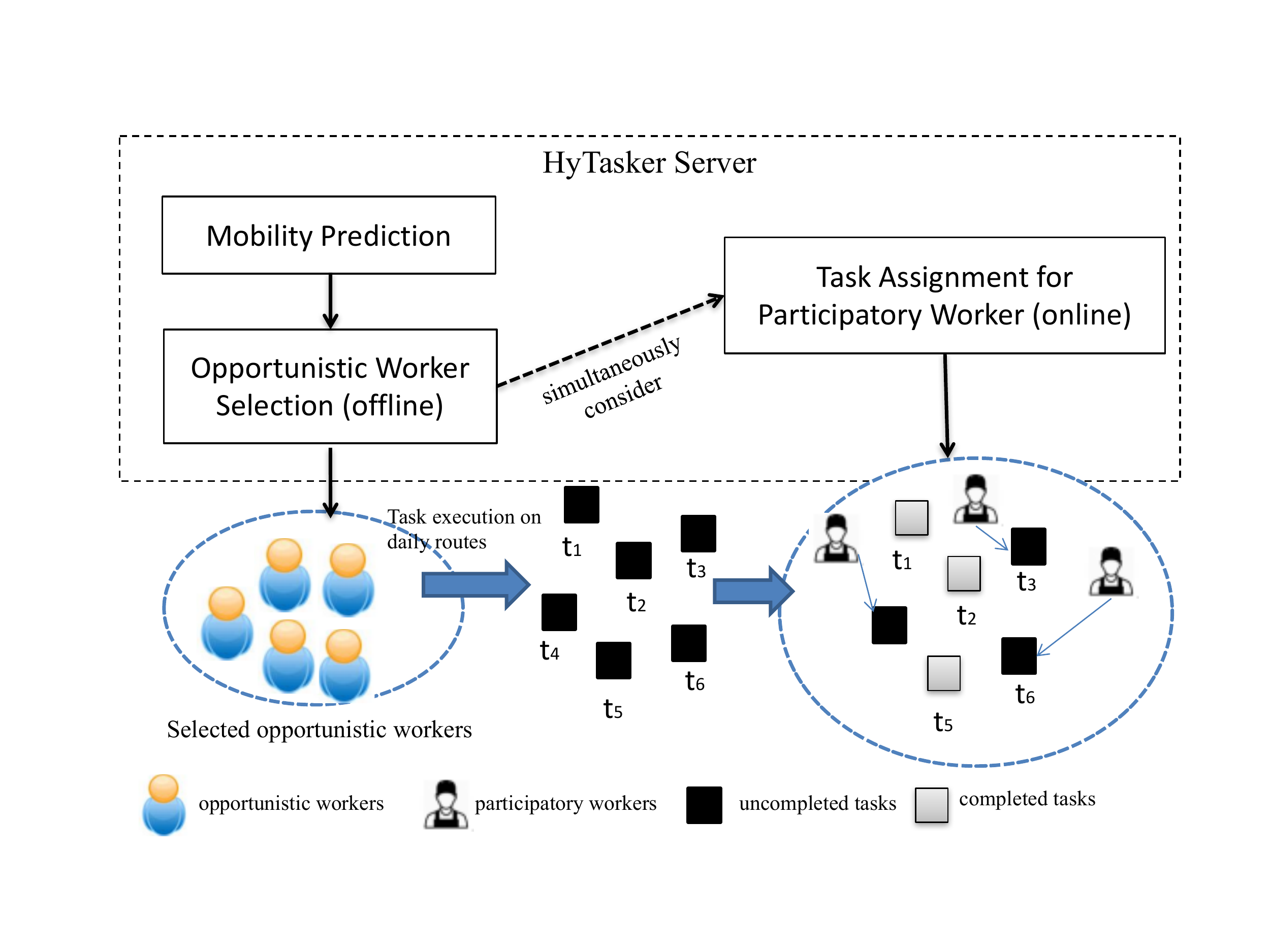}
	
	\caption{The overview design of HyTasker  \label{fig_1}}
\end{figure}

As illustrated in Fig.~\ref{fig_1}, the design of HyTasker mainly includes two phases: \textit{In the offline phase,} it selects a set of workers (called the \textit{opportunistic workers}), and each of them is paid with a fixed reward for the entire sensing period. These workers will complete sensing tasks online during their daily routine. \textit{In the online phase,} HyTasker further spends the rest of the budget by assigning another set of workers (called the \textit{participatory workers}) with certain tasks that have not been completed by the opportunistic workers. The participatory workers will move intentionally to the sensing locations to finish the tasks and get incentive rewards in proportion to their travel distances. The key points of these two phases are summarized as follows, respectively.

\begin{itemize}
	\item \textit{Opportunistic workers selection (offline).} To solve the above-defined hybrid task allocation problem, the biggest challenge lies in the offline selection of the opportunistic workers. Since the opportunistic workers and the participatory workers share a total budget constraint, we cannot consider them separately. Instead, when selecting the opportunistic workers offline, we must consider the future online task assignments for the participatory workers. But the challenge is that, during the offline phase of opportunistic worker selection, we cannot foresee the precise locations of the participatory workers and the completion status of tasks in the online phase. To address this, we propose a heuristic greedy based opportunistic workers selection algorithm by simultaneously considering the online task assignments of the participatory workers.The main process and basic idea of this phase will be presented in Section 3.3, and the detailed algorithms will be illustrated in Section 4.
	\item  \textit{Task assignment for the participatory workers (online).} Note that when selecting the opportunistic workers offline, the task assignments of the participatory workers are predicted actually delivered to the workers. Thus, in the online phase, we will re-do the task assignments based on real-time location information reported by the candidate participatory workers and the actual task completion status. The goal is to cover locations where the tasks are not completed by the opportunistic workers. Since the online task assignment problem for the participatory workers is well studied in \cite{33,14}, we directly adopt the maximum-flow based algorithms in \cite{33,14} and focus the remaining text on the offline phase above.
	
\end{itemize}

\subsection{Opportunistic Worker Selection: Main Process and Basic Idea}
\begin{figure}[!t]
	
	\centering \includegraphics[width=.5\textwidth]{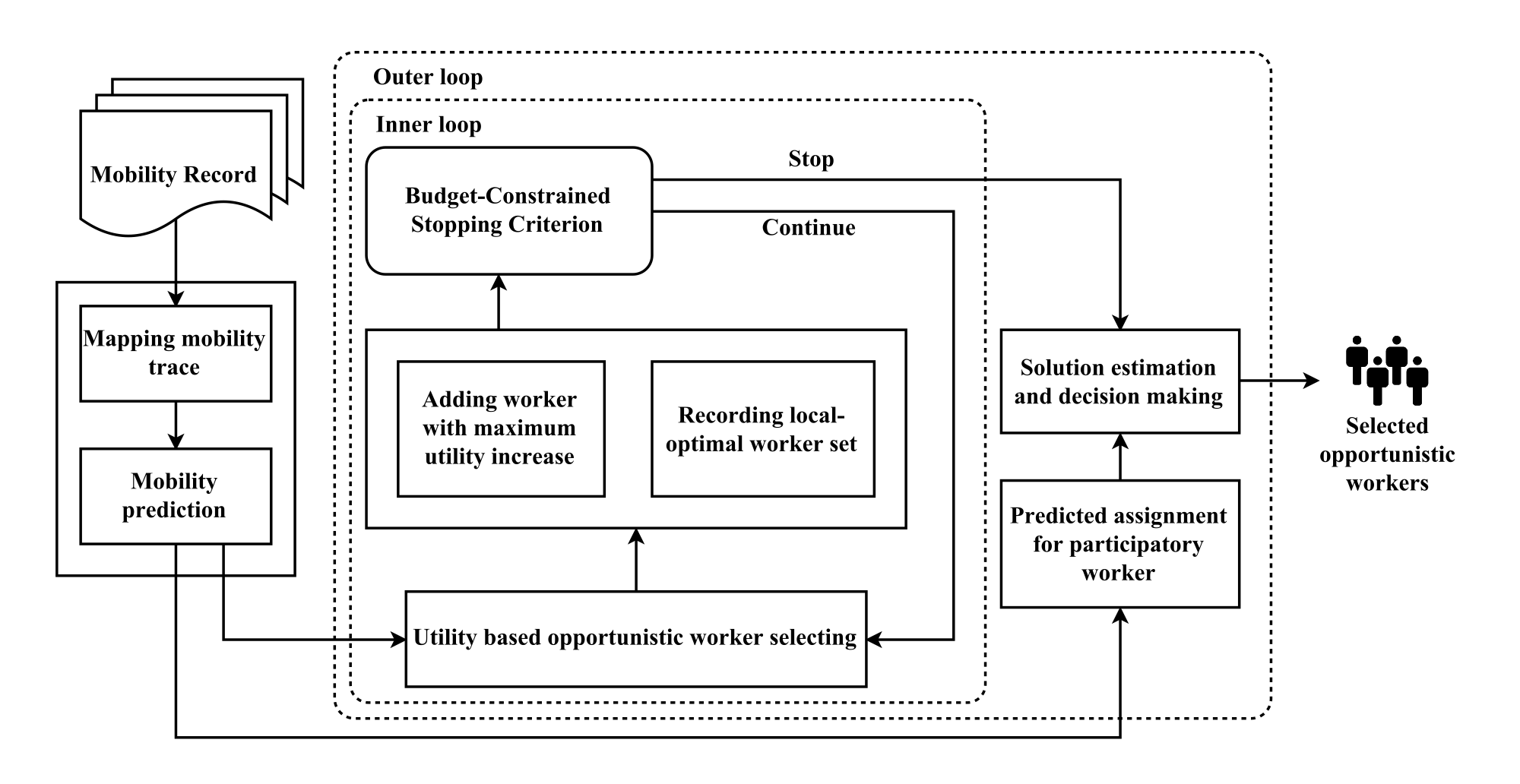}
	
	\caption{Opportunistic worker selection in the offline phase: Main process)  \label{fig_process}}
\end{figure}

The key technical challenge of HyTasker lies in how to do the offline opportunistic worker selection, which requires the joint consideration of the future possible task assignment of the participatory workers. To address this challenge, we propose a nested-loop process (as shown in Fig.~\ref{fig_process}). In this process, the inner-loop searches for and determines which opportunistic worker should be selected, while the outer-loop process determines the best set of workers by estimating the number of completed tasks for a given subset of opportunistic workers and predicting the task assignments of the participatory workers. Finally, the output is the subset of the opportunistic workers that achieves the maximum estimated number of completed tasks.

The proposed process above is more technically challenging and complicated than the state-of-the-art studies for opportunistic worker selection (such as \cite{21,24,30,29}) in the following two aspects.

\begin{figure}[!t]
	
	\centering \includegraphics[width=.5\textwidth]{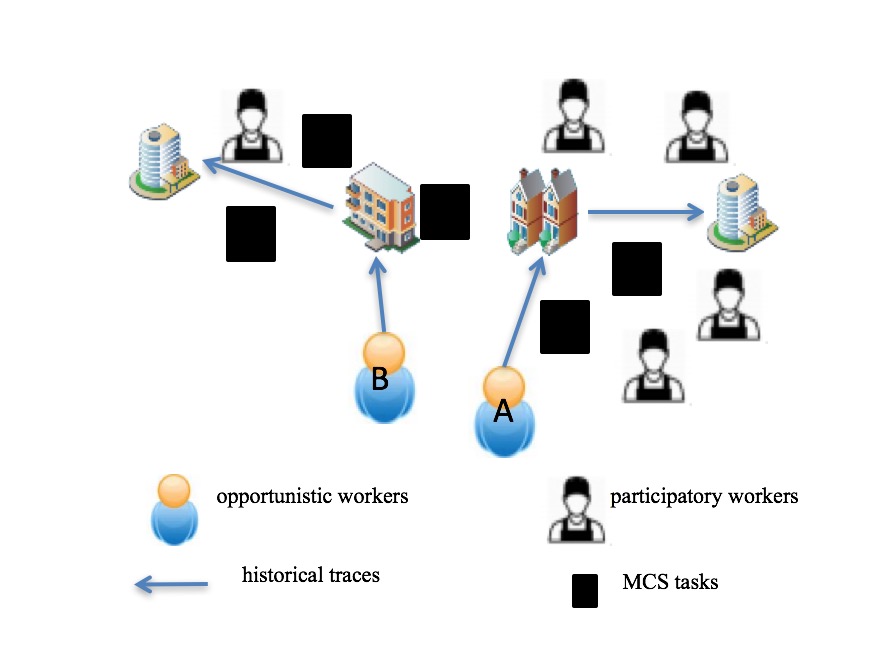}
	
	\caption{An example to illustrate the priority of opportunistic workers (We assign higher priority to worker B than A, because B is more likely to complete tasks in areas where the participatory workers are sparsely distributed)  \label{fig_2}}
\end{figure}

First, for the inner loop, the challenge is how to select the most beneficial opportunistic workers. Existing studies incrementally select workers with the highest estimated coverage gain \cite{21,24,30,29}. However, HyTasker should further consider the future online task assignments of the participatory workers. Specifically, this inner-loop process is designed based on a key idea, that is, we prefer to select the opportunistic workers with two characteristics: (1) workers who can complete more tasks; (2) workers who can complete tasks located in areas where the participatory workers are sparsely distributed according to their historical mobility records. As illustrated by the example in Fig.~\ref{fig_2}, we assign higher priority to the opportunistic worker B than A, because B is more likely to complete tasks within areas with fewer participatory workers. In this paper, we introduce the concept of location entropy \cite{40} in social network community to realize this idea, which will be described in Section  4.2 with more details.

Second, for the existing work \cite{21,24,30,29}, the worker selection process ends when the total budget has been used up, and the finally obtained set of workers is the output. However, in our problem, it is not optimal if we spend all the budget on the opportunistic workers. Therefore, after adding and selecting one worker, we record the obtained subset of opportunistic workers as a snapshot. Ultimately, we determine which subset (snapshot) should be selected by simultaneously considering the predicted task assignments of the participatory workers online. This component will be illustrated with more details in section 4.3.

%% file: core.tex
\section{CORE SUPPORTING ALGORITHMS}
To implement the opportunistic worker selection process in Fig.~\ref{fig_process}, we need to further design three supporting algorithms. First, we need to predict the mobility of all candidate workers (Section 4.1). Second, for the inner loop, we should determine which worker should be selected (Section 4.2). Third, as the execution of each iteration forms a feasible solution (one snapshot), we should further determine which one is optimal (Section 4.3). Finally, we present the algorithm complexity analysis in Section 4.4.

\subsection{Mobility Profiling and Prediction}
Similar to \cite{24,30,31}, this paper assumes that the sensing range of a task is within a cell tower's connection, and the opportunistic workers will complete tasks when they connect to the cell tower. Thus, this algorithm predicts the probability of each worker connecting to different towers at least once during the sensing period. We count the average number of connections by each worker $w_u$ at each cell tower $c_i$, which is denoted as $\lambda_{u,i}$. For example, we set the entire sensing period as one day. To estimate $\lambda_{u,i}$ for of a specific day, we count the average number of connections by $w_u$ at $c_i$ during each day in the historical mobility and connection records. Assuming that the connection sequence follows an inhomogeneous Poisson process \cite{24,30,31}, the probability of worker  $w_u$ connecting to cell tower $c_i$ for $h$ times during a specific day can be modeled as:
\begin{equation}
\varphi_{u,i}(h)=\lambda_{u,i}^{h}*e^{-\lambda_{u,i}}/h!
\end{equation}

Therefore, we can estimate the probability of worker $w_u$ connecting at least once during a day at $c_i$ as follows:
\begin{equation}
Pro_{u,i}=\sum_{h=1}^{\infty}\varphi_{u,i}(h)=1-e^{-\lambda_{u,i}}
\end{equation}

Thus we predict the probability of a candidate opportunistic worker  $w_u$ completing task at $c_i$ as:
\begin{equation}
\alpha_i(w_u) = 1-e^{-\lambda_{u,i}}
\end{equation}

\subsection{Utility-based Candidate Solution Generation}
HyTasker iteratively selects the most beneficial opportunistic worker, and the pseudocode of this process is presented in Algorithm~\ref{alg:OWS}. First of all, HyTasker prefers to select opportunistic workers who will visit as many task locations (subareas) as possible, which is similar to traditional worker selection studies. Besides this criteria, with the hybrid task allocation paradigm that we propose, HyTasker also prefers to select opportunistic workers who will visit locations that are far from the participatory workers, because this reduces the traveling cost of online task allocation for the participatory workers. For example, if two opportunistic workers are predicted to visit the same number of locations, we will prefer the one who can visit locations where the participatory workers are sparsely distributed. 

Thus, HyTasker assigns priority to each task location by considering the past visits of participatory workers. Intuitively, the larger the number of visits to a task location, the lower the priority we assign to that task location. One na\"{\i}ve method is to set the priority of each task location as inversely proportional to the total number of historical visits of all participatory workers. However, this na\"{\i}ve measurement neglects the distribution of visits among different participatory workers, which should also be considered. The reason is that, if the visits belong to a small set of users, it is a bit risky for the online task allocation to count on this small proportion of frequently-visiting users. During the online phase, these users may be too far away or even decline to accept tasks. In summary, we should assign higher priority to task locations with fewer total visits and more concentrated (less uniform) distribution of visits.  Based on the above observation, we introduce the concept of \textit{location entropy}  to characterize the priority of each task location, which considers both the total number and the distribution of the visits among participatory workers. A location would have higher entropy (lower priority) if there are many visits and the visits are distributed more evenly among participatory workers. In contrast, a location will have lower entropy (higher priority) if there are fewer total visits or the distribution of the visits is restricted to only a few participatory workers.

The location entropy of a task $t_i$ is defined as follows:
\begin{equation}
Entropy(t_i)= -\sum_{pw\in PW_i}\frac{|Count_{pw,t_i}|}{|Count_{t_i}|}\times log\frac{|Count_{pw,t_i}|}{|Count_{t_i}|}
\end{equation}
where $PW_i$ denotes the set of participatory workers visited the tower which $t_i$ belongs to, $Count_{t_i}$ denotes the times of the tower which $t_i$ belongs to was visited by $PW_i$, and $Count_{pw,t_i}$ denotes the times of worker $pw$ visited the tower which $t_i$ belonged to.

Inspired by this concept, we formally define the priority of each task based on its location entropy as follows:
\begin{equation}
Weight(t_i) = \frac{1/Entropy(t_i)}{\sum_{t\in T}(1/Entropy(t))}
\end{equation}

Then, the utility increase of adding one candidate opportunistic worker is calculated as follows:
\begin{equation}
\begin{aligned}
&Utility(OW_f\cup\{ow_j\}) - Utility(OW_f) = \\ 
& \sum_{t_i\in T}Weight(t_i)\times \varPhi(i,OW_f\cup\{ow_j\}) \\
& - \sum_{t_i\in T}Weight(t_i)\times \varPhi(i,OW_f)
\end{aligned}
\end{equation}
where $\varPhi(j,OW_f)$ will be illustrated with more details in Eq.(9).

\algrenewcommand{\algorithmiccomment}[1]{\hfill$\blacktriangleright$ #1}
\renewcommand{\algorithmicrequire}{\textbf{Input:}}
\renewcommand{\algorithmicensure}{\textbf{Output:}}
\begin{algorithm}[htb]         
	\caption{Utility-based candidate solution generation}             
	\label{alg:OWS}                  
	\begin{algorithmic}[1]                
		\Require candidate opportunistic workers $OW$; candidate participatory workers $PW$; total budget constraint $B$.
		\Ensure the set of candidate opportunistic worker set $SS$.
		
		\State set $OW_f^{'}=\varnothing$
		\While {$|OW_f^{'}|*I_c<B$}
		\State set $MaxUtility=0$
		\For{each $ow_j\in OW/OW_f^{'}$}
		\If{$Utility(OW_f^{'}\cup \{ow_j\})>MaxUtility$}
		\State $MaxUtility=Utility(OW_f^{'}\cup \{ow_j\})$ \Comment{$Utility$ is be defined in Eq.(8)}
		\State $BestW=ow_j$
		\EndIf
		\EndFor
		
		\State $OW_f^{'} = OW_f^{'} \cup \{ow_j\}$
		\State $OW=OW-\{ow_j\}$
		\State append $OW_f^{'}$ to $SS$
		\EndWhile	
		
		\Return $SS$ 
	\end{algorithmic}
\end{algorithm}

\subsection{Solution Estimation and Decision Making}
At the end of each iteration, we record a snapshot, i.e., a subset of candidate opportunistic workers(see line 12 in Algorithm~\ref{alg:OWS}). Since we assume that each selected opportunistic worker will be given the same reward, the total number of iterations will be $\lfloor\frac{B}{I_c}\rfloor$. In other words, we get $\lfloor\frac{B}{I_c}\rfloor$ possible solutions for the opportunistic worker selection. We can then compare the optimality of those $\lfloor\frac{B}{I_c}\rfloor$ solutions based on the following steps:  

First, we predict the probability of each task to be completed by a given subset of opportunistic workers. In Eq.(5), we have already obtained the probability of an opportunistic worker completing a given task. Thus, the probability that a task $t_i$ can be completed by a set of selected opportunistic workers $OW_f$ is defined as:
\begin{equation}
\varPhi(i,OW_f) = 1-\prod_{ow_i\in OW_f}(1-\alpha_i(ow_i))
\end{equation}

\begin{figure}[!t]
	
	\centering \includegraphics[width=.5\textwidth]{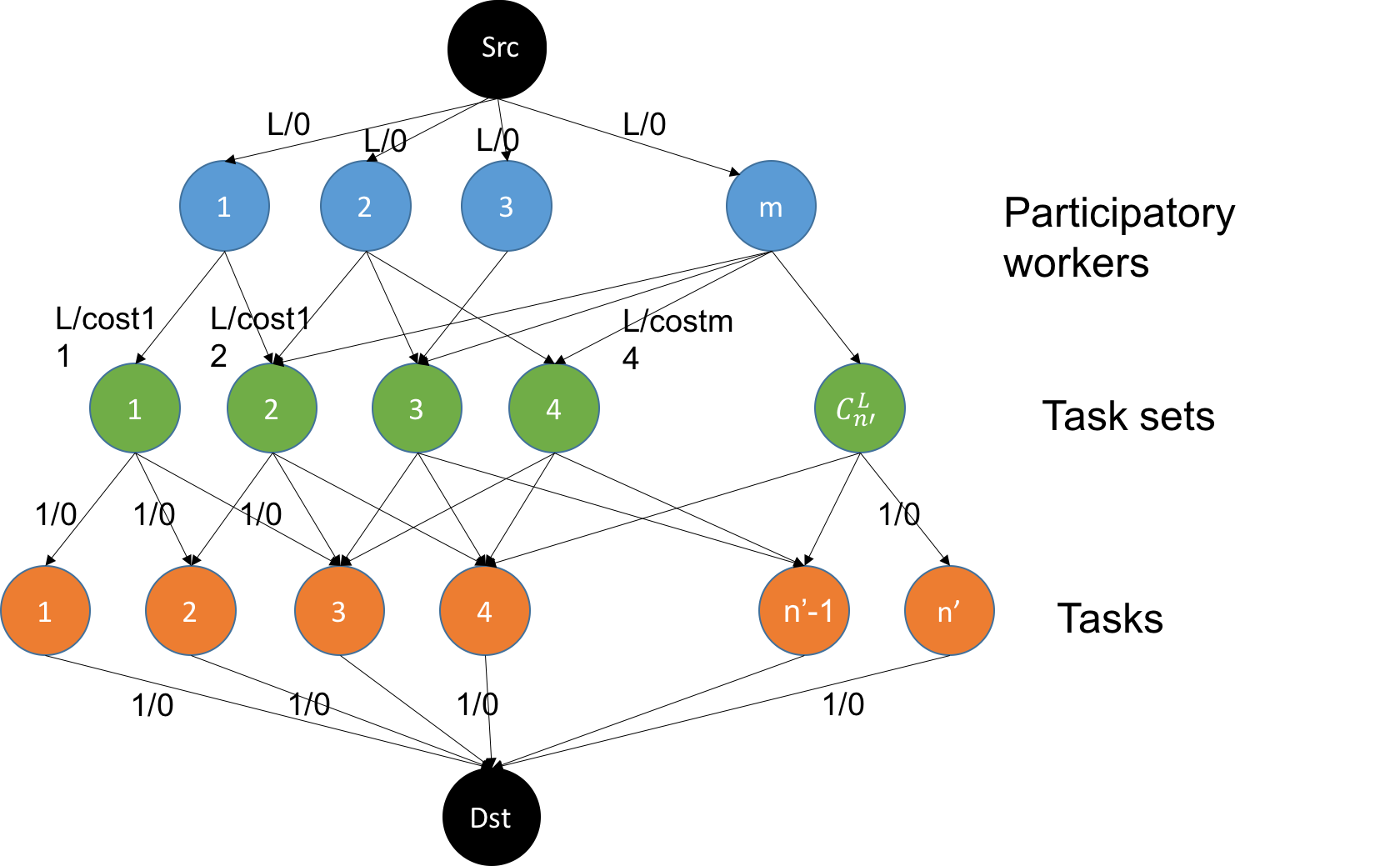}
	
	\caption{The structure of maximum-flow network in a certain round of Monte-Carlo simulations  \label{fig_maxflow}}
\end{figure}

Second, we predict the optimal task assignments of the participatory workers by leveraging the maximum-flow-based algorithms \cite{14,33}. In \cite{14,33}, the assumption is that the locations of workers are known and no task is completed. In contrast, in the offline worker selection phase of HyTasker, we only know the probability of workers' locations and the task completion status. Hence, we adopt multiple rounds of Monte-Carlo simulations to address this issue. Specifically, in each round of simulation, we first generate the completion status of each task and the location of each participatory worker based on the probability. Then, we construct the network of the maximum-flow based on the generated worker location, task completion status, and the  workers' acceptance rate, and sequentially adopt the maximum-flow-based algorithm in \cite{14} to perform task assignment. The average number of completed tasks for multiple rounds of simulations is taken as the final result. Fig.~\ref{fig_maxflow} shows the network structure of the maximum-flow based algorithm. This network contains three levels of nodes: participatory workers, task sets, and tasks, and the edges linking different nodes are associated with specific capacity and cost of the flow.  On the edges linking the blue and green nodes, the cost is the total distance of the shortest path for a specific participatory worker to complete a certain task set. For details about how the shortest path is obtained and how the maximum flow algorithm is executed, interested readers can refer to \cite{14}. The pseudocode of the above predicted task assignment of participatory workers is presented in Algorithm~\ref{alg:PWS}.

\algrenewcommand{\algorithmiccomment}[1]{\hfill$\blacktriangleright$ #1}
\renewcommand{\algorithmicrequire}{\textbf{Input:}}
\renewcommand{\algorithmicensure}{\textbf{Output:}}
\begin{algorithm}[htb]         
	\caption{Task Assignment of Participatory Workers (offline predicate)}             
	\label{alg:PWS}                  
	\begin{algorithmic}[1]                
		\Require candidate participatory workers $PW$; budget constraint $B_p$; tasks $T$; selected opportunistic workers $OW$
		\Ensure the selected subset $V_f$; number of complete tasks $TW$
		\State calculate probability of $t_i\in T$ completed by $OW$
		\State $CompletedTasks=0$
		\For{i=0;i<$Rounds$;i++}
		\State simulate to generate the completed task $CT\subseteq T$
		\State simulate to generate participatory workers' location
		\State simulate if $pw_k\in PW$  is wiling to accept tasks based on the acceptance rate $ac_k$.
		\State select $C_{|CT|}^{L}$ task-task sets from task set $T-CT$
		\State calculate the shortest paths and travel cost
		\State construct the flow network $G=(V,E,C,W)$
		\State initialize flow $f$ to 0
		
		\While there exists an augmenting path in the residual network $G_f$
		\State select the augmenting path $p^{*}$ with minimum cost 
		\State $c_f(p^{*}) = L$
		\State augment flow $f$ along $p^{*}$ with $c_f(p^{*})$
		\EndWhile
		\State $CompletedTasks=CompletedTasks+f+|CT|$
		\EndFor

		\Return $CompletedTasks/Rounds$
	\end{algorithmic}
\end{algorithm}

Third, we estimate the total number of completed tasks by both the opportunistic workers and the participatory workers. The subset of opportunistic workers with maximum estimated total number of completed tasks will be selected as the final output of the offline phase.

\subsection{Algorithm Complexity Analysis}
In this section, we analyze the time complexity of the proposed opportunistic worker selection algorithm. The inner-loop process needs to estimate the utility increase of all unselected opportunistic workers can make. The running time complexity of the inner-loop process of each iteration will be $O(|T|\times |OW|)$. After each iteration selects the best opportunistic worker, the task assignment of the participatory workers will consume $O(|PW|\times L\times (|PW|+|T|+C_{|T|}^{L}))$\cite{14}. The outer-loop process will be run $\lfloor\frac{B}{I_c}\rfloor$ times, so the time complexity is $O(\lfloor\frac{B}{I_c}\rfloor\times (|T|\times |OW| + |PW|\times L\times (|PW|+|T|+C_{|T|}^{L})))$. 

%% file: eval.tex
\section{EXPERIMENTAL EVALUATION}

\subsection{Experimental Purposes and Baselines}
The goal of our experiments is to compare the performance of HyTasker and other baseline methods under different situations, such as different number of tasks, different number of workers, different total incentive budget, and so on. The performance comparison metrics include the number of completed tasks and running time.  

Specifically, we provide the following baseline task allocation methods for comparative studies.

\textbf{OPP} \textit{(opportunistic mode based approach)}: This algorithm only uses opportunistic mode to maximize the number of completed tasks while keeping the budget constraint. It spends all budget to select the opportunistic workers. Similar to \cite{24}, OPP iteratively selects workers offline with the maximum utility increase until the total budget is used up, and the selected workers will complete tasks online during their daily routines. This baseline method is used to test whether the hybrid approach is more effective than the pure opportunistic mode approach.

\textbf{PAR} \textit{(participatory mode based approach)}: This algorithm only uses participatory mode to maximize the number of completed tasks while keeping the budget constraint. PAR spends all budgets to allocate the participatory workers online, and the workers will intentionally move to the task locations. Specifically, it adopts the maximum-flow based algorithm in \cite{33} to select the optimal set of task-and-worker pairs. This baseline is designed to evaluate whether the hybrid approach is better than the pure participatory mode approach.

\textbf{BP-Hybrid} \textit{(budget partition based hybrid approach)}: This algorithm tries to divide the budget into different proportions, and then directly adopts the state-of-the-art task allocation methods for opportunistic mode and participatory mode, respectively. For each round, it first randomly generates the proportion of the budget division. Then, it uses the greedy algorithm in \cite{24} to select the opportunistic workers and adopts the maximum-flow based algorithm in \cite{33} to perform online task assignment for the participatory workers. BP-Hybrid repeats the above process for $\lfloor\frac{B}{I_c}\rfloor$ rounds, and the maximum number of completed tasks achieved is chosen as the final result.


\subsection{Datasets and Experimental Setups}
\begin{figure}
	
	\centering \includegraphics[width=.5\textwidth]{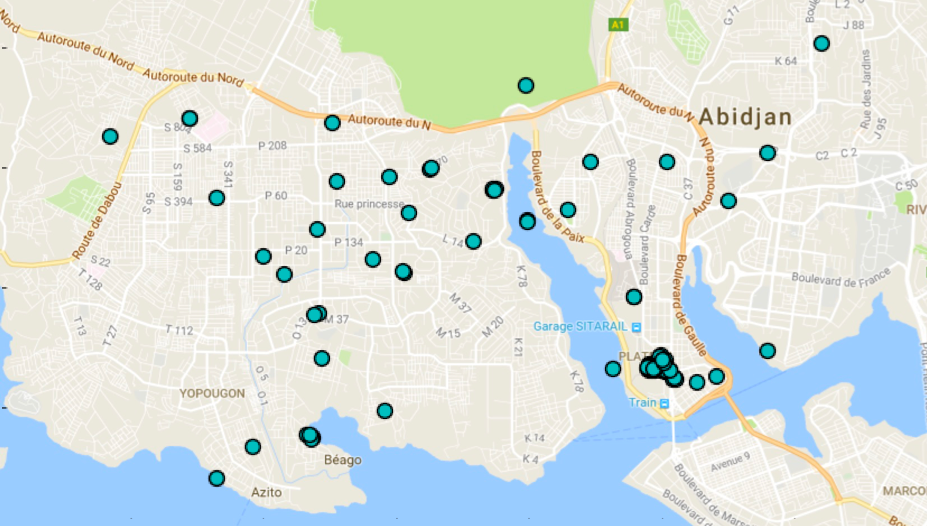}
	
	\caption{Entire sensing area and the distribution of cell towers in our experiment  \label{fig_map}}
\end{figure}

The dataset we used in evaluation is the D4D dataset \cite{37}, which contains two types of data records in Ivory Coast. One contains the information about cell towers, including tower id, latitude and longitude. The other one contains 50,000 users' phone call records. We select users randomly every 2 weeks (for weekdays) with anonymized ids and in total 10 sets of ten-day period of records are stored in the dataset. Here, for each set of ten-day records, we use the first nine-day records to model users' mobility patterns (described in Section 4.1), and use the 10th day as the test sensing period to execute the online task assignment algorithm and evaluate the number of completed tasks. Specifically, we extracted records of the downtown area (100 cell towers with 1000 mobile users), as shown in Fig.~\ref{fig_map}.  We need to set a number of parameters in the experiments, which are divided into task-relevant and worker-relevant as follows.  

For the tasks, their locations are randomly distributed to a group of cell towers within the target area, and there may be several tasks located in the same cell tower. In the experiments, the sensing range of each task is within its deployed cell tower's connection range. Similar to \cite{24,30,31}, the selected opportunistic workers can complete tasks in piggyback manner \cite{27} when they connect to the cell towers. Moreover, each task is assumed to last for one day from 8:00am to 6:00pm and can be completed at any time within this period on the 10th day. Here, to simplify the problem, we assume that the online task assignment for the participatory workers is executed one hour before the end of the sensing period (i.e., 5:00pm on the 10th day), and the participatory workers can complete all assigned tasks before the end of the sensing period.

For the workers, the settings are different for the opportunistic workers and the participatory workers. Each mobile user in the D4D dataset is set as either a candidate opportunistic worker or a candidate participatory worker with a certain probability (i.e., opportunistic worker:$\gamma$, participatory worker: $1-\gamma$).The opportunistic workers are selected from candidates who make phone calls near these cell towers. For the participatory workers, the maximum number of tasks they can perform is randomly set to be between 2 to 5. Their initial locations, when the online task assignment is performed, are set as the most frequently-visited cell towers in their historical mobility records. Moreover, the spatial region of each participatory worker is set as a rectangular region bounded by his/her historically-visited areas. Similar to \cite{45,46}, we also simulate the acceptance rate of each participatory worker with a Gaussian distribution (with mean value $\mu$), and further test the performance of different approaches by varying $\mu$.

The reward for each opportunistic worker is set to 10 US dollars for the entire sensing period (i.e., one day), while the reward per kilometer for the participatory workers is set to 10 US dollars. Here we use the Manhattan Distance \cite{41} to measure the travel distance between two locations. 

The aforementioned parameter settings are summarized in Table~\ref{table_2}. We carried out the experiments using a laptop computer with an Intel Core i7-4710HQ Quad-Core CPU and 16GB memory. HyTasker and other baseline methods were implemented with the Java SE platform on a Java HotSpotTM 64-Bit Server.

\begin{table}
	\centering
	\caption{Summary of Experimental Parameter Settings \label{table_2}}
	\begin{tabular}{|c|c|}
		\hline
		\textbf{Parameters} & \textbf{Settings}\\
		\hline
		Total budget & 200,400,600,800,1000 \\
		\hline
		Number of workers & 100,200,300,400,500 \\
		\hline
		Number of tasks & 30,60,90,120,150  \\
		\hline
		Percentage of OW ($\gamma$) & 0.1,0.2,...,0.9 \\
		\hline
		Mean value of acceptance rate $\mu$ & 0.2,0.4,0.6,0.8,1.0 \\
		\hline
		Maximum number of tasks for PW & Random generate between 2 and 5 \\
		\hline
		reward per kilometer for PW & 10 \\
		\hline
		reward for each opportunistic worker & 10 \\
		\hline
	\end{tabular}
\end{table}

\subsection{Experimental Results}
In this section, we first report the experimental results of different methods and compare their performance with regard to the number of completed tasks and running time under various settings. Here, as the online task assignment algorithm of HyTasker and BP-Hybrid is the same, the "running time" in this section refers to the program execution time in the offline opportunistic worker selection phase (note that PAR does not have the offline worker selection phase).

\subsubsection{Different value of total budget}
\begin{figure}
	\subfigure
	\centering 
	\includegraphics[width=1.75in]{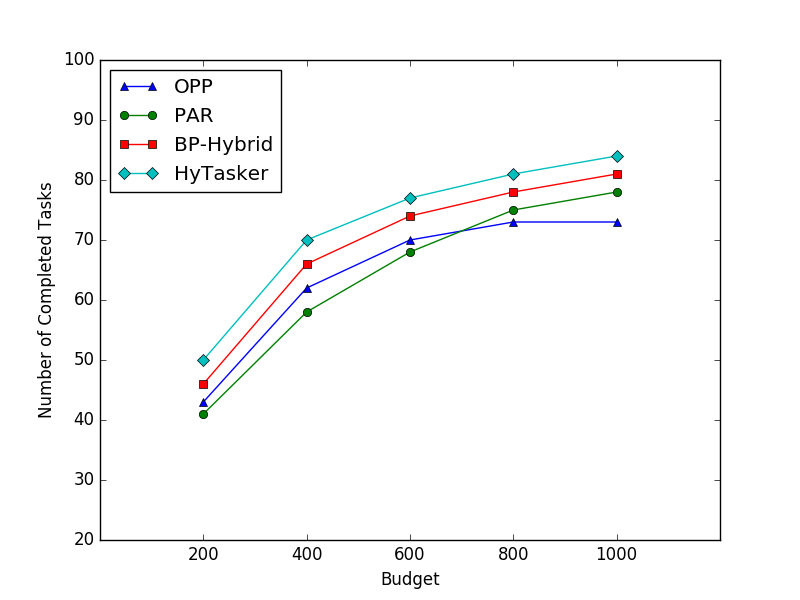}
	\includegraphics[width=1.75in]{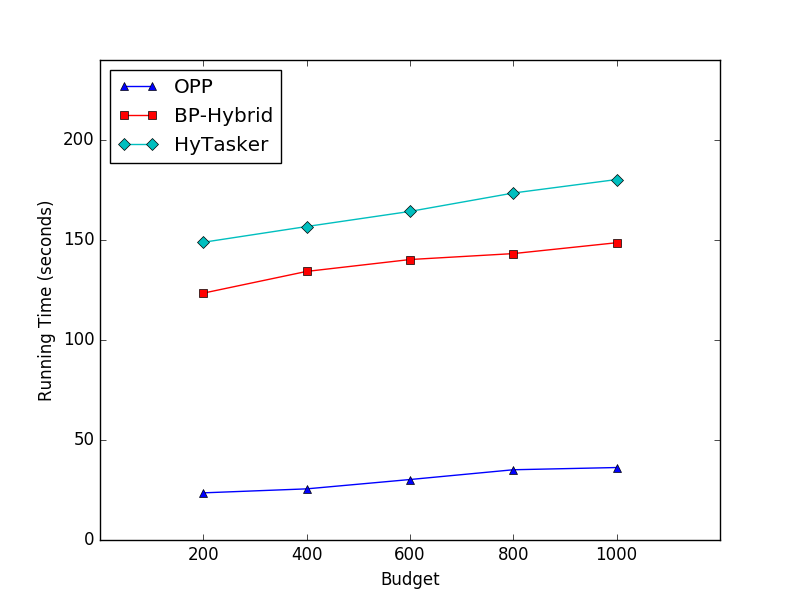}
	\subfigure
	
	\caption{Performance comparison under different settings of total budget \label{fig_budget}}
\end{figure}

In Fig.~\ref{fig_budget},  we compare the performance of different methods under various settings of the total budget. In order to control other variables, we fix the number of tasks at 90, the number of workers at 300, and the value of $\gamma$ at 0.6. Here, we assume that the participatory workers will always accept the assigned tasks. From Fig.~\ref{fig_budget} (left), we can see that the number of completed tasks increases with the total incentive budget for all methods, because a higher budget allows more workers to be recruited to complete more tasks. For the number of completed tasks, HyTasker outperforms other baseline methods in all budget settings. Fig.~\ref{fig_budget} (right) reports the running time of HyTasker, OPP and BP-Hybrid. All results are measured on reasonably efficient implementation of the various algorithms. Although HyTasker needs longer running time than BP-Hybrid and OPP, its running time is less than 5 minutes for various settings of the total budget. Since the algorithm is executed offline and the experiments were run on a laptop computer, this computation time is acceptable. To implement a real-world MCS system, shorter computation time can be achieved by using parallel algorithms or deploying HyTasker on a more powerful commercial server.

\subsubsection{Different number of tasks}
\begin{figure}
	\subfigure
	\centering 
	\includegraphics[width=1.75in]{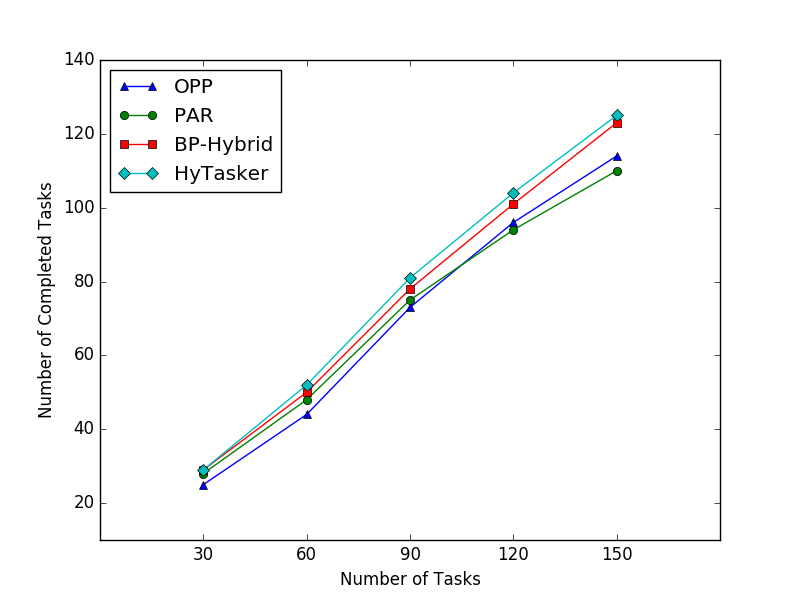}
	\includegraphics[width=1.75in]{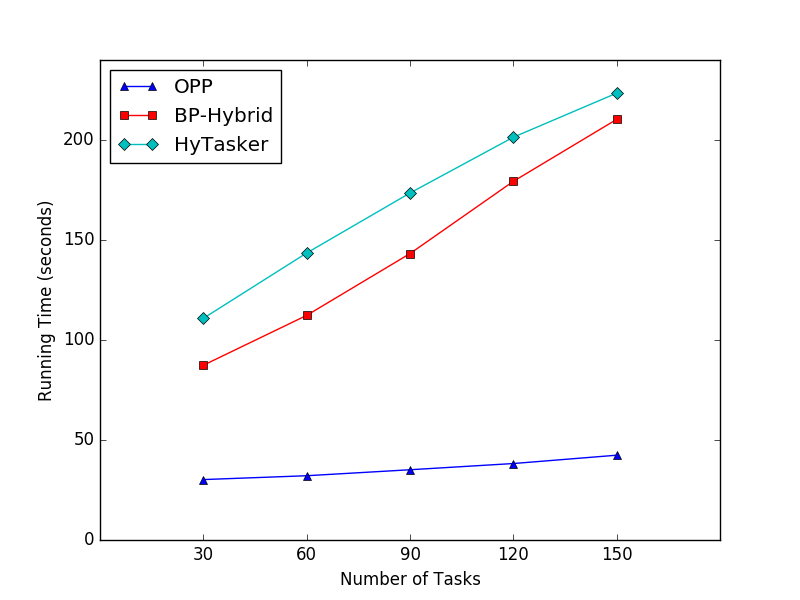}
	\subfigure
	
	\caption{Performance comparison under different number of tasks \label{fig_task}}
\end{figure}

In Fig.~\ref{fig_task}, we compare the performance of different methods under different number of tasks. Here we fix the total incentive budget at 800 US dollars, the number of workers at 300, and the value of $\gamma$ at 0.6. Here, we assume that the participatory workers will always accept the assigned tasks. From Fig.~\ref{fig_task} (left), we can see that HyTasker outperforms other baseline methods in all settings of the number of tasks. Fig.~\ref{fig_task} (right) reports the running time of different methods. Again, although HyTasker needs longer running time than OPP and BP-Hybrid, its running time is less than 5 minutes for different number of tasks. Since the algorithm is executed offline, the computation time is acceptable.

\subsubsection{Different number of workers}
\begin{figure}
	\subfigure
	\centering 
	\includegraphics[width=1.75in]{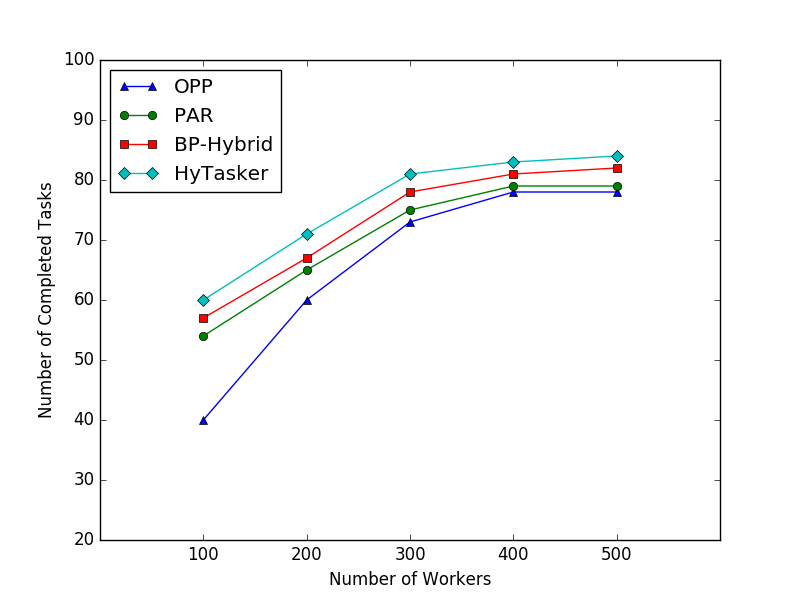}
	\includegraphics[width=1.75in]{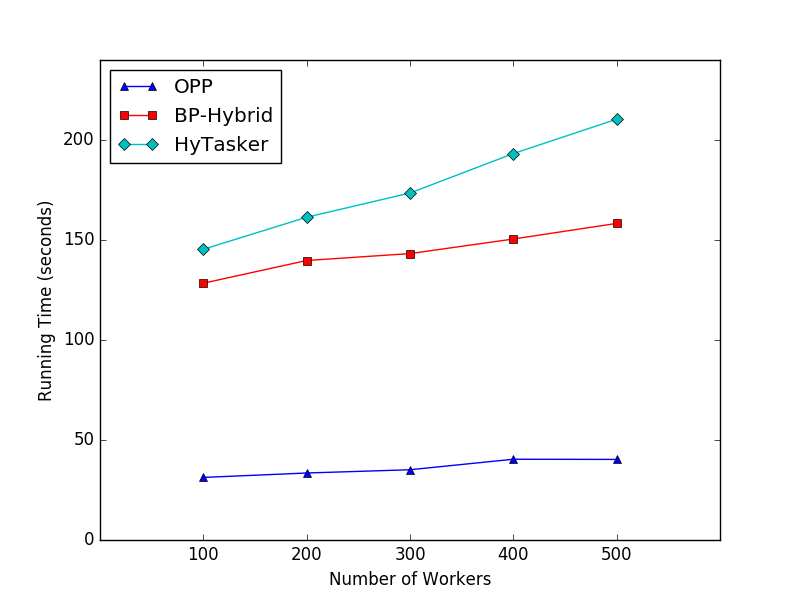}
	\subfigure
	
	\caption{Performance comparison under various number of workers \label{fig_work}}
\end{figure}

In Fig.~\ref{fig_work}, we present the performance comparison under different number of workers. Here we fix the total incentive budget at 800 US dollars, the number of tasks at 90, and the value of $\gamma$ at 0.6. From Fig.~\ref{fig_work} (left), we can see that the number of completed tasks increase when there are more candidate workers for all methods, because the task allocation algorithms have more candidate workers to choose from and can generate a better allocation plan. We can also see that HyTasker outperforms other baseline methods in all settings of the number of workers. From Fig.~\ref{fig_work} (right), we can also see that HyTasker needs longer running time in offline worker selection than OPP and BP-Hybrid, but the computation time is acceptable as it is executed offline.

\subsubsection{Different value of \texorpdfstring{$\gamma$}{Lg}}
\begin{figure}
	\subfigure
	\centering 
	\includegraphics[width=1.75in]{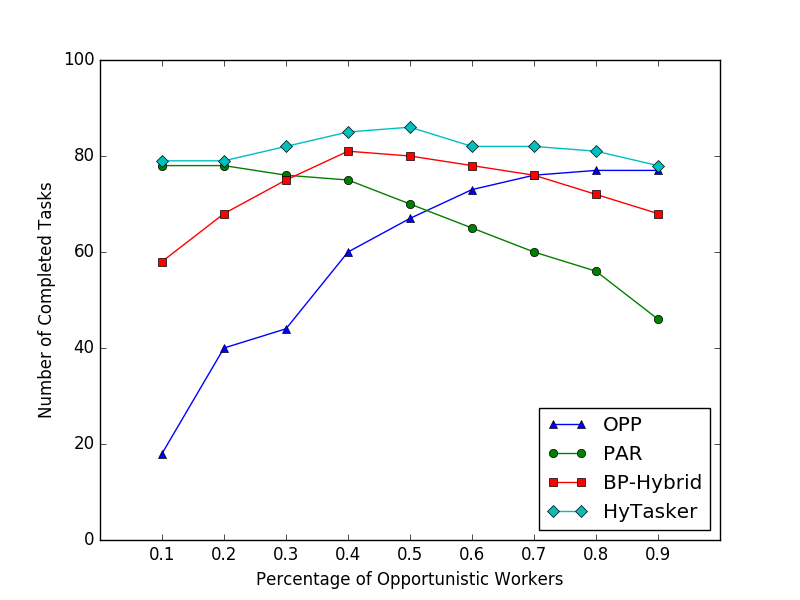}
	\includegraphics[width=1.75in]{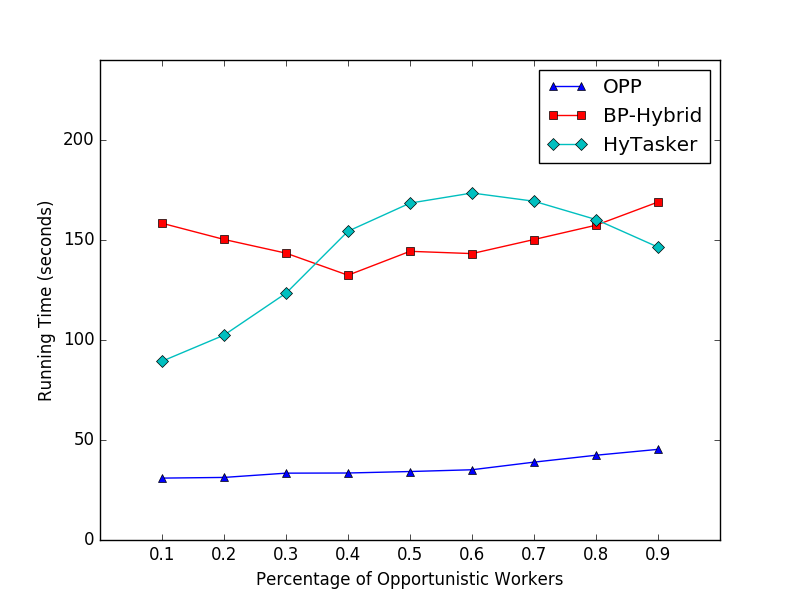}
	\subfigure
	
	\caption{Performance comparison under various percentage of opportunistic workers \label{fig_gamma}}
\end{figure}

In Fig.~\ref{fig_gamma}, we illustrate the performance comparison under different values of $\gamma$. Here we fix the total incentive budget at 800 US dollars, the number of tasks at 90, and the number of workers at 300. We assume that the participatory workers will always accept the assigned tasks. From Fig.~\ref{fig_gamma} (left), we can also see that HyTasker outperforms other baseline methods in all settings. We can also see that with the increase of $\gamma$, OPP can achieve a better performance for the number of completed tasks, while the performance of PAR becomes worse. This is because, with the fixed total number of candidate workers and increase of $\gamma$, there are more candidate opportunistic workers to choose from, while the number of candidate participatory workers becomes fewer. From Fig.~\ref{fig_gamma} (right), we can also see that HyTasker needs longer running time in offline worker selection, but it is less than 5 minutes under all settings.

\subsubsection{Different value of \texorpdfstring{$\mu$}{Lg}}
\begin{figure}
	\subfigure
	\centering 
	\includegraphics[width=1.75in]{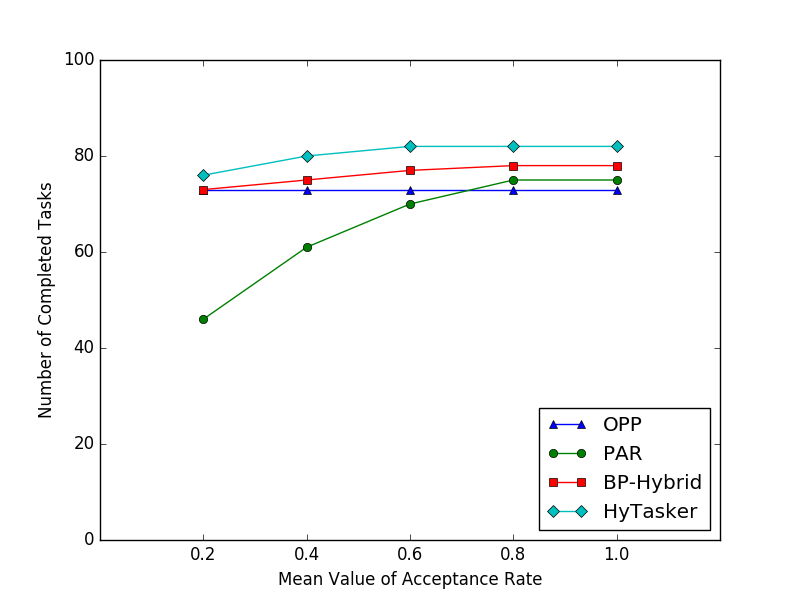}
	\includegraphics[width=1.75in]{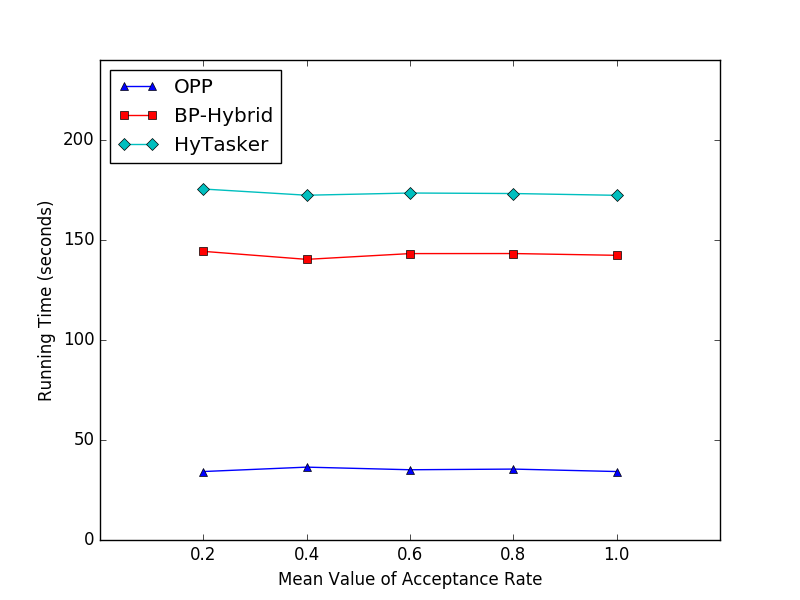}
	\subfigure
	
	\caption{Performance comparison under different values of $\mu$ (the mean value of task acceptance rate of $PW$) \label{fig_mu}}
\end{figure}

Similar to \cite{45,46}, we assume that the acceptance rate of each participatory worker follows a Gaussian distribution with a mean value of $\mu$, and present the performance of different approaches in Fig.~\ref{fig_mu} by varying $\mu$. Here we fix the total incentive budget at 800 US dollars, the number of tasks at 90, the number of workers at 300, and the value of $\gamma$ at 0.6. From Fig.~\ref{fig_mu} (left), we can also see that HyTasker consistently outperforms other methods in all settings of $\mu$. According to Fig.~\ref{fig_mu} (right), $\mu$ has almost no impact on the running time of HyTasker. 

\subsubsection{Detailed Analysis and Implications}
The experimental results above (from Fig.~\ref{fig_budget} to Fig.~\ref{fig_gamma}) show an overall comparison of different methods under various parameter settings. In this subsection, we will further present some details and corresponding analysis, which can verify some of our observations and intuitions.

\begin{figure}
	\subfigure
	\centering 
	\includegraphics[width=1.75in]{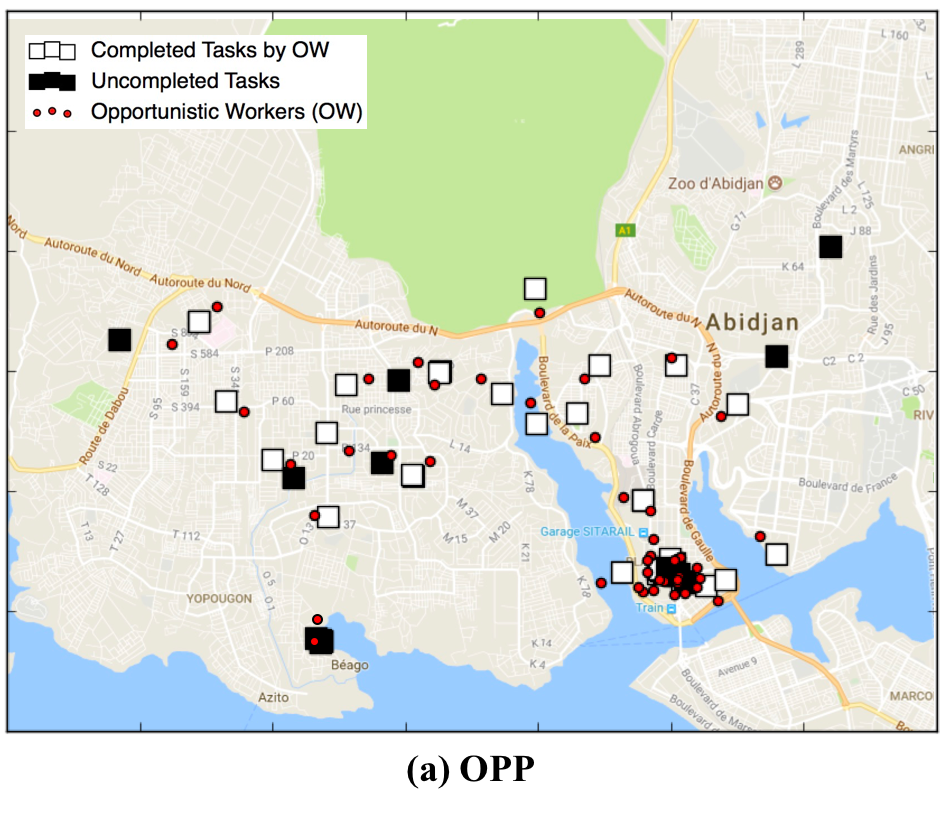}
	\includegraphics[width=1.75in]{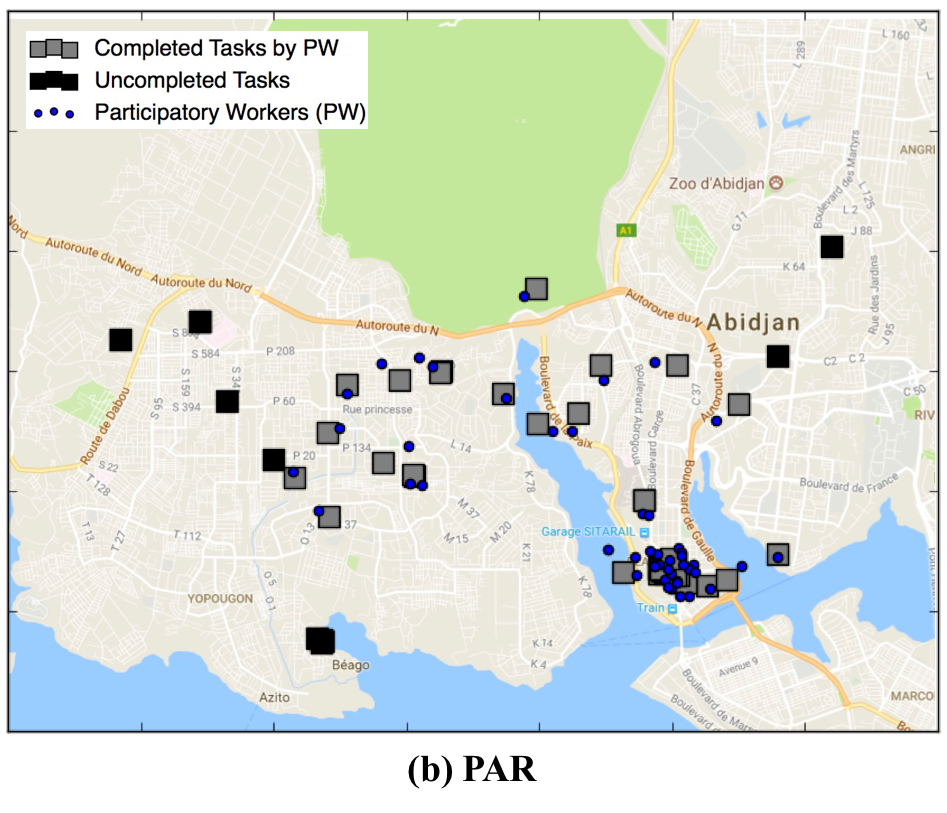}
	\subfigure
	
	\subfigure
	\centering 
	\includegraphics[width=1.75in]{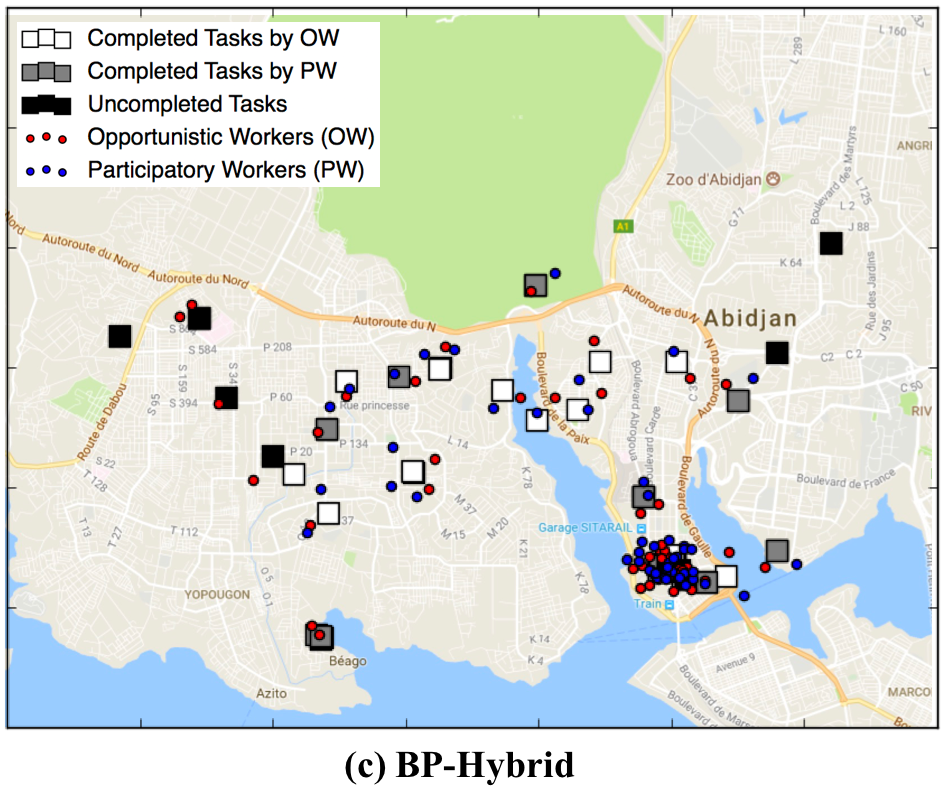}
	\includegraphics[width=1.75in]{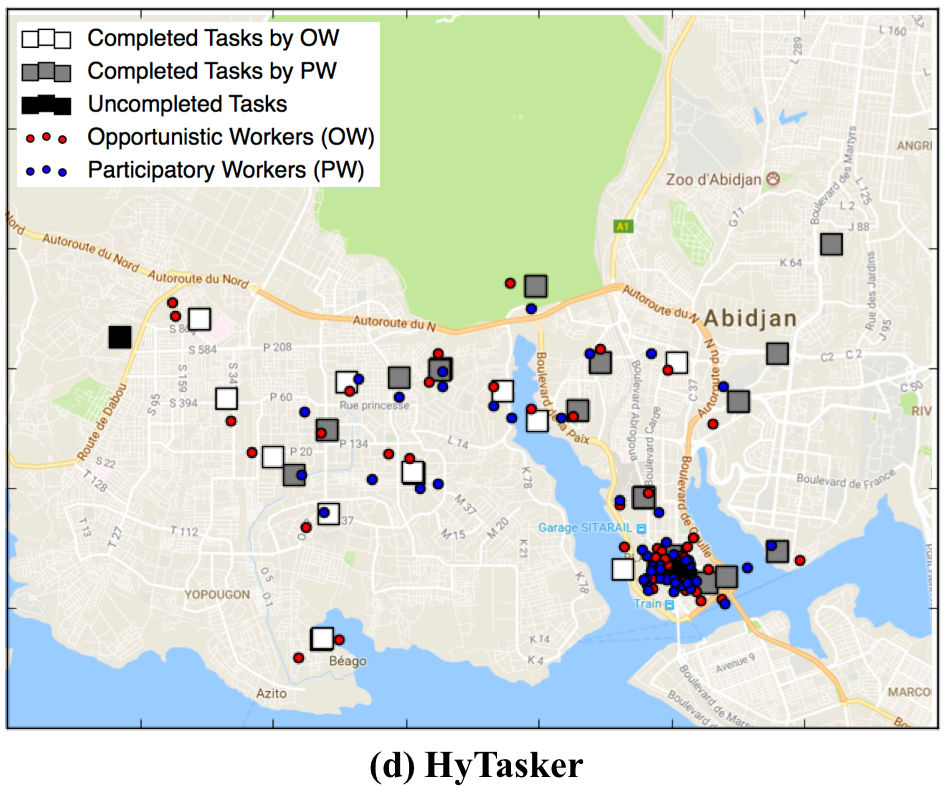}
	\subfigure
	
	
	\caption{Worker distribution and task completion status for a single round:  (a) OPP; (b) PAR; (c) BP-Hybrid; (d) HyTasker. \label{fig_round}}
\end{figure}

Fig.~\ref{fig_round} (a)-(d) visualize the distribution of tasks, locations of workers and final task completion status under a specific setting (i.e., number of workers is 300, number of tasks is 60, total budget is 800, acceptance rate $\mu$ is 1.0, and $\gamma$ is 0.6). For the completed tasks achieved by BP-Hybrid and HyTasker, we further use different legends to show whether they are completed by participatory workers or opportunistic workers.

From Fig.~\ref{fig_round}(a) and (d), we can see that compared with OPP, the advantage of HyTasker is that it can complete some tasks in the worker-sparse areas, thus the number of completed tasks is increased. The comparison of Fig.~\ref{fig_round}(c) and Fig.~\ref{fig_round}(d) further demonstrates why HyTasker outperforms BP-Hybrid. From the distribution visualization of two types of workers, we can see that compared with BP-Hybrid, HyTasker can better leverage the opportunistic workers to complete tasks in areas where participatory workers are sparsely distributed. This indicates that: 1) the joint optimization of offline and online phase is more beneficial than considering them separately, and 2) it is effective to consider the density of participatory workers when selecting the opportunistic workers.

\begin{figure}
	
	\centering \includegraphics[width=.5\textwidth]{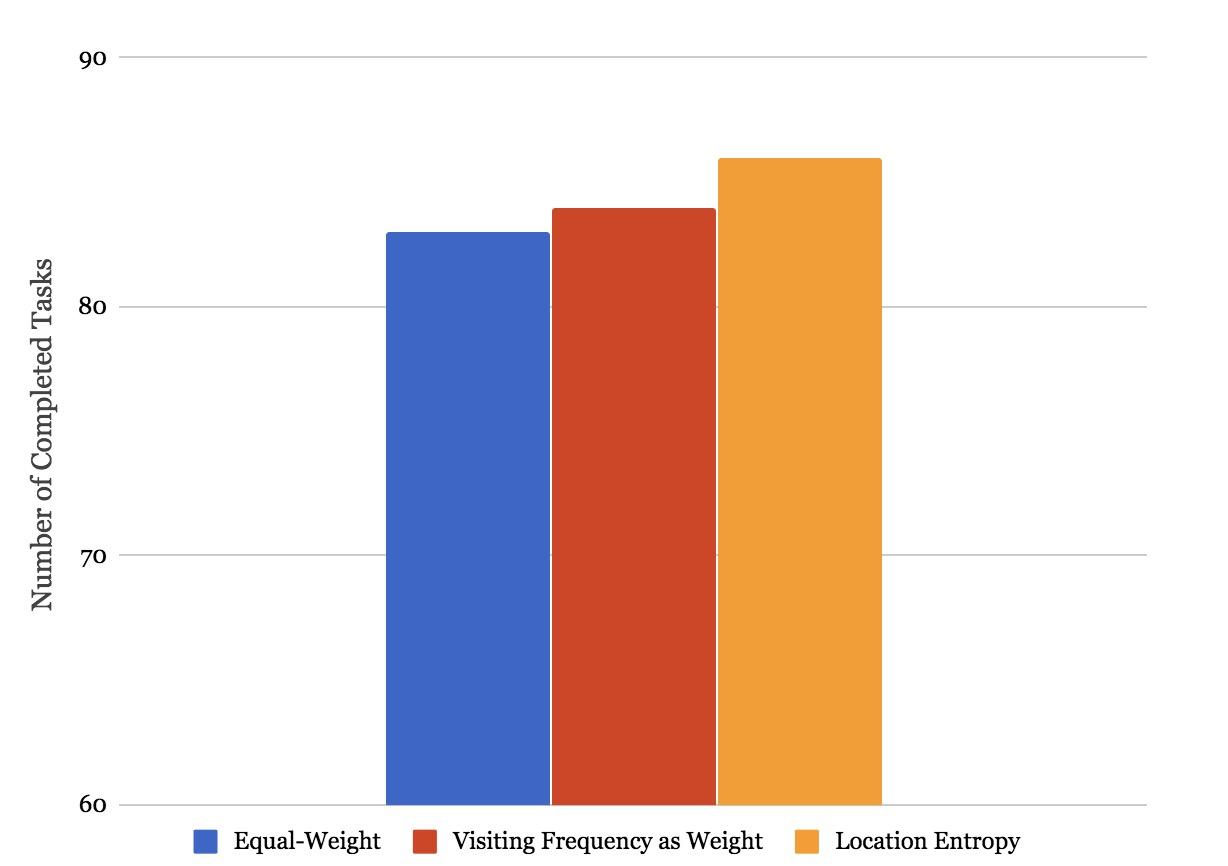}
	
	\caption{Demonstrating the effectiveness of location entropy by comparing HyTasker with other two variants.  \label{fig_last}}
\end{figure}

In addition, to test whether the use of location entropy is beneficial, we further compare HyTasker with two variants of HyTasker without using location entropy. One variant (called "Equal-Weight") sets equal weight for each task location, and the other variant (called "Visiting Frequency as Weight") sets the priority to be inversely proportional to the total number of past visits. Fig.~\ref{fig_last} presents the average number of completed tasks among multiple rounds of experiments when the total number of tasks is fixed at 90. We can see that with the adoption of location entropy, HyTasker can complete more tasks than the two variants.

	
	